\documentclass[a4paper,11pt]{article}
\pdfoutput=1 

\usepackage{jheppub} 
\usepackage{hyperref}
\usepackage{caption}
\usepackage{subcaption}
\usepackage[T1]{fontenc} 
\usepackage{listings}

\usepackage{tikz}
\usetikzlibrary{matrix,arrows,snakes,shapes,decorations.fractals,decorations.pathmorphing,patterns,decorations.markings}

\usepackage{upgreek}
\usepackage{slashed}

\newcommand{\mc}{\mathcal}
\usepackage{mathrsfs}

\usepackage{tikz}

\title{Bootstrapping line defects in AdS$_3$/CFT$_2$} 
\author[a]{Gabriel Bliard}
\author[b]{Diego H. Correa}
\author[b,c]{Mart\'in Lagares}
\author[b]{Ignacio Salazar Landea}

\affiliation[a]{Laboratoire de Physique, \'Ecole Normale Sup\'erieure, Universit{\'e} PSL, CNRS, Sorbonne Universit{\'e}, Universit{\'e} Paris Cit{\'e}, 
24 rue Lhomond, F-75005 Paris, France}
\affiliation[b]{Instituto de Fisica La Plata, Universidad Nacional de La Plata,
C.C. 67, 1900 La Plata, Argentina}
\affiliation[c]{Max-Planck-Institut f\"{u}r Physik, Werner-Heisenberg-Institut, Boltzmannstr. 8, 85748 Garching, Germany}

\emailAdd{gabriel.bliard@ens.fr, correa@fisica.unlp.edu.ar, martinlagares95@gmail.com, peznacho@gmail.com}

\preprint{}

\abstract{

We study correlators of insertions along 1/2 BPS line defects in the holographic dual to type IIB string theory in $AdS_3 \times S^3 \times T^4$ with mixed Ramond-Ramond and Neveu Schwarz-Neveu Schwarz three-form flux. 
These defects break the symmetries of the bulk CFT$_2$ as $PSU(1,1|2)^2\times SO(4) \rightarrow PSU(1,1|2)\times SU(2)$, defining displacement and tilt supermultiplets.
We focus on the two-, three- and four-point functions of these supermultiplets, which we compute using analytic conformal bootstrap up to next-to-leading order in their strong-coupling expansion. We obtain a bootstrap result that only depends on two OPE coefficients. We perform a Witten diagram check of the bootstrap result, obtaining an holographic interpretation of the two OPE coefficients that are not constrained by the bootstrap procedure.

}

\begin{document} 
\maketitle



\section{Introduction}

In recent years there has been a renewed  interest in understanding conformal field theories in the presence of defects \cite{Cardy:1984bb}, which can be realized by imposing boundary conditions on the theory \cite{Cardy:2004hm}, or by inserting extended operators \cite{Drukker:2006xg,Giombi:2017cqn,Billo:2016cpy,Gaiotto:2014kfa}. Among conformal defects, a canonical example is given by Wilson lines, that can be interpreted as heavy charged particles in the vacuum of a gauge theory. Many quantities of physical interest can be obtained from Wilson line correlators, such as the quark-antiquark potential \cite{Maldacena:1998im} or the bremmstrahlung function \cite{Correa:2012at}. 

The existence of a very precise and well-studied holographic interpretation of Wilson lines in the context of AdS/CFT dualities \cite{Maldacena:1997re,Maldacena:1998im} allows for both weak- and strong-coupling analysis of these defects in many theories. In this framework, correlators of insertions along the contour of a Wilson line conformal defect can be obtained at strong coupling in terms of a series of AdS$_2$ Witten diagrams  \cite{Giombi:2017cqn}. However, a bottleneck step in the pursue of this strong-coupling prescription relies in the fact that computing Witten diagrams at subleading orders becomes quickly unfeasible. Many non-perturbative techniques, such as integrability and supersymmetric localization, have been profoundly studied and developed with the goal of overcoming the aforementioned problem \cite{Drukker:2006xg,Drukker:2012de,Correa:2012hh,Giombi:2018qox,Correa:2023lsm,Pestun:2007rz}. In particular, the analytic conformal bootstrap has arisen during the last years as a very useful tool to study the strong-coupling regime of one-dimensional defect conformal field theories (dCFTs) defined by Wilson lines \cite{Liendo:2018ukf,Bianchi:2020hsz,Ferrero:2021bsb,Ferrero:2023gnu,Ferrero:2023znz}. 
These techniques, which are part of the conformal bootstrap program that was reignited in the last years by a series of seminal papers \cite{Rattazzi:2008pe,El-Showk:2012cjh}, consist on the analytic computation of CFT correlators by imposing symmetry constraints and other consistency conditions.

As expected, the bootstrap program becomes increasingly powerful as symmetry is enhanced, which is the case when one considers dCFTs with supersymmetry, i.e. superconformal defect field theories. Indeed, this has been demonstrated in the analytic bootstrap of 1/2 BPS conformal lines in the ${\cal N}=4$ super Yang-Mills (SYM) \cite{Liendo:2018ukf,Ferrero:2021bsb,Ferrero:2023gnu,Ferrero:2023znz} and ABJM \cite{Bianchi:2020hsz} theories. In the former case the defect is left invariant under 8 supercharges, while in the latter the supersymmetry is reduced to 6 supercharges. It is therefore natural to pose the question of whether is it possible to apply analytic bootstrap techniques to study superconformal line defects with even fewer supersymmetries. In this context, we will present in this paper an analytic bootstrap study of 1/2 BPS line defects in the CFT$_2$ which is dual to type IIB string theory in $AdS_3\times S^3\times T^4$ with mixed Ramond-Ramond (R-R) and Neveu Schwarz-Neveu Schwarz (NS-NS) three-form flux. This string theory is obtained in the $S^3 \times S^1 \rightarrow T^4$ limit of string theory in $AdS_3 \times S^3 \times S^3 \times S^1$, and therefore the 1/2 BPS line defects that we will consider in this paper can be obtained as a limit of the ones that were introduced in \cite{Correa:2021sky}. At the level of the symmetry algebra of the defects, this limit implies $\mathfrak{d}(2,1;\sin^2 \Omega) \rightarrow \mathfrak{psu}(1,1|2) \times \mathfrak{su}(2)_A$, where $0<\Omega<\frac{\pi}{2}$ is a parameter that measures the relative size of the two $S^3$ spheres in the $AdS_3 \times S^3 \times S^3 \times S^1$ background and the $A$ subindex is used to note that the $\mathfrak{su}(2)_A$ factor is an automorphism of the $\mathfrak{psu}(1,1|2)$ algebra. The corresponding dCFTs are then invariant under only 4 supercharges, providing therefore an interesting set up to study analytic bootstrap techniques in defects with fewer supersymmetries than the cases of the 1/2 BPS lines in ${\cal N}=4$ SYM and ABJM\footnote{Another recent example where a line invariant under 4 supercharges is studied with bootstrap techniques is presented in \cite{Pozzi:2024xnu}. For other examples of the application of analytic bootstrap methods see for instance \cite{Gimenez-Grau:2019hez}.}. Moreover, the interpolation between pure R-R and pure NS-NS backgrounds introduces an additional coupling to the theory, which enriches the bootstrap analysis. 

In defect superconformal field theories there are supermultiplets that naturally arise from the breaking of the symmetries of the bulk theory due to the presence of the defect. For the 1/2 BPS case that we will study, the bulk symmetry is broken as $PSU(1,1|2)^2 \times SO(4) \rightarrow PSU(1,1|2) \times SU(2)_A$, where the $SO(4)$ and $SU(2)_A$ factors represent automorphisms of the $PSU(1,1|2)^2$ and $PSU(1,1|2)$ superconformal groups, respectively. Therefore, these defects define a \textit{displacement} supermultiplet that arises from the symmetry breaking $PSU(1,1|2)^2 \rightarrow PSU(1,1|2) $, and a \textit{tilt} supermultiplet that is defined due to the breaking $SO(4) \rightarrow SU(2)_A$. In this paper we have focused on the analytic bootstrap computation of two-, three- and four-point functions of fields within each of those two supermultiplets and up to next-to-leading order in their strong-coupling expansion. We have found that supersymmetry completely determines the two- and three-point functions up to a single constant for each supermultiplet. As for the four-point functions, supersymmetry allowed us to determine all four-point correlators between insertions of the tilt of displacement supermultiplets in terms of three constants and four functions of a single cross-ratio. We have computed these constants and functions using analytic conformal bootstrap, obtaining a final result that only depends on two coefficients. This allowed us to obtain the next-to-leading order CFT data associated to the line defects, expressed in terms of the two parameters left undetermined by the bootstrap.
This aligns well with the intuition from the string theory dual perspective, where correlators also depend on two parameters: the 't Hooft coupling and the parameter governing the R-R and NS-NS flux mixing. We provide a Witten diagram calculation of the four-point correlators that precisely match their cross-ratio dependence, offering a holographic interpretation of the two coefficients that parametrize the bootstrap result.

The paper is organized as follows. We begin with a discussion of the symmetry algebra of the defects and the corresponding representation theory in Section \ref{sec algebra and reprs}. There we introduce the displacement and tilt supermultiplets and discuss the topological sector of their correlators. Then, in Section \ref{sec 2-pt and 3-pt} we move to a computation of the two- and three-point functions of those supermultiplets. In Section \ref{sec 4-pt} we present the analysis of the four-point functions. We begin the section with an analysis of the constraints imposed by supersymmetry on those correlators. Then, after introducing the OPE expansion of the four-point functions, we present the analytic bootstrap computation of those correlators, discussing all the different constraints that are imposed in the procedure. In Section \ref{Sec: Holographic description} we perform an holographic check of the bootstrap results. Finally, we give our conclusions in Section \ref{sec conclusions}. We include an Appendix with useful formulas that complement the holographic analysis of Section \ref{Sec: Holographic description}.

\section{Symmetry algebra and representation theory}
\label{sec algebra and reprs}

As introduced above, in the following we will focus on a 1/2 BPS line defect defined on the CFT$_2$ dual to type IIB string theory in $AdS_3 \times S^3 \times T^4$ with mixed R-R and NS-NS flux. Therefore, we find convenient to begin with an analysis of the supersymmetry algebra of the defect and of its corresponding representations. 

\subsection{Algebra}

\label{susy algebra defect}

We will start the discussion by presenting the supersymmetry algebra of the dCFT$_1$ described by the 1/2 BPS line operator. The symmetries of the $AdS_3 \times S^3 \times T^4$ background form a $\mathfrak{psu}(1,1|2)^2\times \mathfrak{so}(4) \times \mathfrak{u}(1)$ algebra \cite{Borsato:2013qpa}, where the extra $\mathfrak{so}(4)\times \mathfrak{u}(1)$ acts as an automorphism. For the 1/2 BPS line defect the corresponding symmetry algebra is $\mathfrak{psu}(1,1|2)\times \mathfrak{su}(2)_A$. The bosonic $\mathfrak{su}(1,1) \times \mathfrak{su}(2)_R$ subalgebra of the $\mathfrak{psu}(1,1|2)$ describes the conformal and R-symmetries of the defect, while the extra $\mathfrak{su}(2)_A$ is again an automorphism. Let us note that in the literature it is common to preserve only the $\mathfrak{u}(1)$ generated by the Cartan of the $\mathfrak{su}(2)_A$ automorphism\footnote{An interesting discussion on this subject is given in Section 2.2 of \cite{Beisert:2006qh}.}. The non-vanishing commutators of this algebra \cite{Borsato:2013qpa} are
\begin{align}
[D, P] &= P, 
& [D, K] &= -K,  & [P, K] &= -2 D, 
\nonumber \\
[R^a_b, R^c_d] &= \delta^c_b R^a_d-\delta^a_d R^c_b,
&[R^a_b, Q_{c\dot{c}}]  &= \delta^a_c Q_{b \dot{c}}-\frac{1}{2} \delta^a_{b} Q_{c \dot{c}}, & [R^a_b, S_{c\dot{c}}]  &= \delta^a_c S_{b \dot{c}}-\frac{1}{2} \delta^a_{b} S_{c \dot{c}},  \nonumber \\
[L^{\dot{a}}_{\dot{b}}, L^{\dot{c}}_{\dot{d}}] &= \delta^{\dot{c}}_{\dot{b}} L^{\dot{a}}_{\dot{d}}-\delta^{\dot{a}}_{\dot{d}} L^{\dot{c}}_{\dot{b}}, 
&[L^{\dot{a}}_{\dot{b}}, Q_{c\dot{c}}]  &= \delta^{\dot{a}}_{\dot{c}} Q_{c \dot{b}}-\frac{1}{2} \delta^{\dot{a}}_{\dot{b}} Q_{c \dot{c}}, 
&[L^{\dot{a}}_{\dot{b}}, S_{c\dot{c}}]  &= \delta^{\dot{a}}_{\dot{c}} S_{c \dot{b}}-\frac{1}{2} \delta^{\dot{a}}_{\dot{b}} S_{c \dot{c}}, 
\nonumber \\
[D, Q_{a\dot{a}}] &= \frac{1}{2} Q_{a\dot{a}},  & [D, S_{a\dot{a}}] &= -\frac{1}{2} S_{a\dot{a}},
& \{ Q_{a\dot{a}}, Q_{b\dot{b}}\} 
&=-\epsilon_{ab} \epsilon_{\dot{a}\dot{b}} P, \nonumber \\
[P,  S_{a\dot{a}}] &= -Q_{a\dot{a}}, & [K,  Q_{a\dot{a}}] &= S_{a\dot{a}}, & \{ S_{a\dot{a}}, S_{b\dot{b}}\} &=-\epsilon_{ab} \epsilon_{\dot{a}\dot{b}} K, \nonumber 
\\
& & \{ Q_{a\dot{a}}, S_{b\dot{b}}\} &=-\epsilon_{ab} \epsilon_{\dot{a}\dot{b}} D  -  \epsilon_{\dot{a}\dot{b}} R_{ab},    & & 
\label{algebra}
\end{align}
where $P,K$ and $D$ are the generators of the conformal group, $R^a_b$ are the generators of the $\mathfrak{su}(2)_R$ R-symmetry, $L^{\dot{a}}_{\dot{b}}$ are the generators of the $\mathfrak{su}(2)_A$ automorphism, and $Q_{a\dot{a}}$ and $S_{a\dot{a}}$ are the supercharges\footnote{To uplift the $\mathfrak{psu}(1,1|2)\times \mathfrak{su}(2)_A$ algebra to a $\mathfrak{d}(2,1;\sin^2{\Omega})$ algebra (that describes 1/2 BPS line operators in the CFT$_2$ dual to type IIB string theory on AdS$_3\times$S$^3\times$S$^3\times$S$^1$) we just have to modify the $\{ Q_{a\dot{a}}, S_{b\dot{b}}\}$ anticommutator in order for it to be
\begin{equation}
 \{ Q_{a\dot{a}}, S_{b\dot{b}}\} =-\epsilon_{ab} \epsilon_{\dot{a}\dot{b}} D - \sin^2\Omega \,\epsilon_{\dot{a}\dot{b}} R_{ab} - \cos^2\Omega  \,\epsilon_{ab} L_{\dot{a}\dot{b}} \,.
 \label{d21a}
\end{equation}}. Note that the dotted $\mathfrak{su}(2)_A$ and undotted $\mathfrak{su}(2)_R$ indices take values in $\pm$ and $\dot{\pm}$, respectively, and we are using $\epsilon_{+-}=\epsilon_{\dot{+}\dot{-}}=1$ and $R_{ab}=\epsilon_{bc}R_a^c$.

\subsection{Representations}
\label{reps}

In order to bootstrap four-point functions of operators inserted along the defect we need to be exhaustive about all the possible representations that can appear in the corresponding OPE expansions. Therefore, let us study the representations of the $\mathfrak{psu}(1,1|2) \times \mathfrak{su}(2)_A$ symmetry algebra of the defect. To describe the states in those representations we will use the notation $|\Delta,r,\ell \rangle$, with
\begin{align}
   D |\Delta,r,\ell \rangle &= \Delta \,|\Delta,r,\ell \rangle \,, \nonumber \\
   R_0 |\Delta,r,\ell \rangle &= \frac{r}{2} \, |\Delta,r,\ell \rangle \,, \nonumber \\
   L_0 |\Delta,r,\ell \rangle &= \frac{\ell}{2} \, |\Delta,r,\ell \rangle \,,
\end{align}
and where $R_0:=R^+_+=-R^-_-$ and $L_0:=L^+_+=-L^-_-$ are the corresponding Cartan generators of the R-symmetry and automorphism groups.

On the one hand, we have $\textit{long representations}$ ${\cal L}^{\Delta}_{r,\ell}$, which are characterized by the three quantum numbers $\Delta, r$ and $\ell$ of their corresponding superprimary state. These representations are given by 16 conformal primaries, that organize as
\footnote{The multiplicity of the different states in the decomposition can be seen from a character analysis.}
\begin{align}
    [\Delta,&r,\ell]\nonumber \\
    [\Delta+\frac 12,r+1,\ell+1] \qquad [\Delta+\frac 12,r+1,\ell-1] &\qquad [\Delta+\frac 12,r-1,\ell+1] \qquad [\Delta+\frac 12,r-1,\ell-1] \nonumber\\
    [\Delta+1,r\pm2,\ell]\qquad 2 \, [\Delta+1,&r,\ell]\qquad[\Delta+1,r,\ell\pm2]\\
   [\Delta+\frac 32,r+1,\ell+1] \qquad [\Delta+\frac 32,r+1,\ell-1] &\qquad [\Delta+\frac 32,r-1,\ell+1] \qquad [\Delta+\frac 32,r-1,\ell-1] \nonumber\\
    [\Delta+2,&r,\ell] \nonumber
\end{align}
These satisfy a unitarity bound 
\begin{align}
    \Delta \geq \frac r 2.
\end{align}
On the other hand, we have \textit{short representations} that preserve half of the supersymmetries, i.e. that are 1/2 BPS. These multiplets are subjected to the shortening condition\footnote{Naively one could also derive $\Delta=-\frac{r}{2}$ from the algebra. However, this condition is not consistent with the $\Delta \geq 0$ constraint that comes from unitarity.}
\begin{equation}
    \Delta=\frac{r}{2} \,,
\end{equation}
that can be derived from the anti-commutation relations
\begin{align}
\label{qs 1}
    \{Q_{\pm \dot{\pm}}, S_{\mp\dot{\mp}} \} &= -D \pm R_0 \,, \\
\label{qs 2}
    \{Q_{\pm \dot{\mp}}, S_{\mp \dot{\pm}}\} &= D \mp R_0 \,.
\end{align}
Therefore, there are only two independent quantum numbers that characterize the superprimary state of these supermultiplets, which we will choose to be $\Delta$ and $\ell$. Consequently, we will use the notation ${\cal A}^{\Delta}_{\ell}$ and ${\cal B}^{\Delta}_{\ell}$ to denote these representations. The structure of the short multiplets is
\begin{align}
\label{short multiplets} 
    \mc{A}^{\Delta}_{\ell}&:\qquad [\Delta,2\Delta,\ell]\rightarrow [\Delta+\frac 12,2\Delta+1,\ell\pm 1]\rightarrow[\Delta+ 1,2\Delta+2,\ell] \nonumber \\
    \mc{B}^{\Delta}_{\ell}&:\qquad [\Delta,2\Delta,\ell]\rightarrow [\Delta+\frac 12,2\Delta-1,\ell\pm 1]\rightarrow[\Delta+ 1,2\Delta-2,\ell]
\end{align}
For example, we have:
\begin{align}
\label{short multiplets 2}
    &\mc{B}^{\frac 12 }_{1}: [\frac 12,1,1]\rightarrow [1,0,2]\oplus [1,0,0] \nonumber\\
    &\mc{B}^{1}_{0}: [1,2,0]\rightarrow [\frac 32,1,1]\rightarrow[2,0,0] \nonumber\\
    &\mc{B}^{1}_{2}: [1,2,2]\rightarrow [\frac 32,1,1]\oplus [\frac 32,1,3]\rightarrow[2,0,2]
\end{align}

Finally, let us comment about the absence of other short representations. More precisely, we will argue that the symmetry algebra of the defect forbids the existence of 1/4 BPS representations\footnote{This is in contrast with the analysis of \cite{Agmon:2020pde}. The main difference is that in \cite{Agmon:2020pde} the authors consider one-dimensional defects embedded in a three-dimensional bulk space, while our defect is embedded in a two-dimensional space. Therefore, in the former case there is an extra $\mathfrak{u}(1)$ symmetry (describing rotations around the defect) that is absent in our case.}. In order to show this, let us suppose that we have a superprimary state $|\Delta, r, \ell \rangle $ that is annihilated by the supercharge $Q_{+\dot{+}}$, i.e.
\begin{equation}
Q_{+\dot{+}} |\Delta, r, \ell \rangle=0 \,,
\end{equation}
Then, from \eqref{qs 1} we get
\begin{equation}
(D-R_0) |\Delta, r, \ell \rangle=0 \,,
\end{equation}
and therefore, using \eqref{qs 2} we arrive at
\begin{equation}
 \{Q_{+ \dot{-}}, S_{-\dot{+}}\} |\Delta, r, \ell \rangle= S_{-\dot{+}} Q_{+ \dot{-}} |\Delta, r, \ell \rangle=0 \,,
\end{equation}
were in the last step we have used that $|\Delta, r, \ell \rangle$ is a superprimary. Then,
\begin{equation}
|| Q_{+ \dot{-}} |\Delta, r, \ell \rangle ||^2= \langle \Delta, r, \ell|S_{-\dot{+}}Q_{+ \dot{-}} |\Delta, r, \ell \rangle=0 \,,
\end{equation}
and consequently
\begin{equation}
     Q_{+ \dot{-}} |\Delta, r, \ell \rangle=0 \,. \label{Eq:All shorts are 1/2 BPS}
\end{equation}
Therefore, we have shown that each superprimary that is annihilated by the $Q_{+\dot{+}}$ charge is also annihilated by the $Q_{+ \dot{-}}$ charge (and one can similarly prove the reciprocal). Then, each short representation is 1/2 BPS\footnote{Let us note that this statement is not true for the more general $\mathfrak{d}(2,1;\sin^2{\Omega})$ supergroup. Indeed, the displacement supermultiplet is 1/4 BPS for the case of 1/2 BPS line defects in the holographic dual to string theory in $AdS_3 \times S^3 \times S^3 \times S^1$, where the symmetry group of the defect is $\mathfrak{d}(2,1;\sin^2{\Omega})$.}.

\subsection{The displacement and tilt supermultiplets}
\label{sec displ and tilt}

As discussed above, the presence of the defect breaks some of the symmetries of the bulk CFT$_2$. In the following we will be interested in the operators with protected scaling dimensions that are associated to those broken CFT$_2$ symmetries.

The most obvious effect of the presence of a line defect is the breaking of translation invariance, which is described in the Ward identity \cite{McAvity:1993ue}
\begin{equation}
\label{ward translations}
    \partial_\mu T^{\mu x}(t,x)= \delta(x) \rho(t) \,,
\end{equation}
where $\mu=t,x$ and $T^{\mu\nu}$ is the stress-energy tensor of the bulk CFT$_2$. From \eqref{ward translations} we can define the \textit{displacement operator} $\rho$, which is identified with a $|2,0,0 \rangle$ state. Since $\rho$ must have protected quantum numbers, it must fit within a short representation. From \eqref{short multiplets} we see that the only short multiplet that can account for a $|2,0,0 \rangle$ state is $\mc{B}^1_{0}$, which can therefore be defined as the \textit{displacement supermultiplet}. One can relate the remaining operators in this supermultiplet with those associated to the breaking of supersymmetry and R-symmetry. Indeed, the breaking $SO(4) \rightarrow SU(2)_R$  of the R-symmetry implies
\begin{equation}
\label{ward R symm}
    \partial_\mu J^{\mu ab}(t,x)= \delta(x) \phi^{ab}(t) \,,
\end{equation}
where $J^{\mu ab}$ is the broken R-symmetry current and $\phi^{ab}$ is an operator of dimension $\Delta=1$, which we can identify with the superprimary $|1,2,0 \rangle$ of the displacement supermultiplet. Let us emphasize that $\phi^{ab}$ is symmetric in its indices. As for the $\chi^{a\dot{a}}$ fermion described by the $|\tfrac{3}{2},1,1\rangle$ state, it can be similarly associated to the breaking of supersymmetry. Therefore, we see that the displacement supermultiplet is related to the breaking $PSU(1,1|2)^2\to PSU(1,1|2) $ of the spacetime symmetry. The structure of the supermultiplet is 
\begin{align}
\label{displ multiplet structure}
    \mc{B}^{1}_{0}:\qquad [1,2,0]\rightarrow [\frac 32,1, 1]\rightarrow[2,0,0]
\end{align}
and the corresponding supersymmetry transformations are
\begin{align}
    Q_{a\dot{a}} \phi_{bc}&= \frac{1}{2} \epsilon_{ab} \chi_{c\dot{a}} +\frac{1}{2} \epsilon_{ac} \chi_{b\dot{a}} \,, \nonumber\\
    Q_{a\dot{a}} \chi_{b\dot{b}}&=  -\epsilon_{\dot{a}\dot{b}} \partial_t\phi_{ab} + \epsilon_{ab}\epsilon_{\dot{a}\dot{b}} \rho \,, \nonumber\\
    Q_{a\dot{a}} \rho&= \frac{1}{2} \partial_t \chi_{a \dot{a}} \,.
\end{align}

On the other hand, the Ward identity that describes the breaking $SO(4)\times U(1) \rightarrow SU(2)_A$ of the automorphism
\begin{equation}
\label{ward automorphism}
    \partial_\mu W^{\mu \dot{a}\dot{b}}(t,x)= \delta(x) \varphi^{\dot{a}\dot{b}}(t) \,,
\end{equation}
allows us to define a $\varphi^{\dot{a}\dot{b}}$ operator, which is identified with a $|1,0,2 \rangle \oplus |1,0,0 \rangle$ state. Above we are introducing the automorphism currrent $W^{\mu \dot{a} \dot{b}}$. Let us note that $\varphi^{\dot{a}\dot{b}}$ is not symmetric, which comes from the fact that we have a broken $U(2)$ subgroup of the automorphism. We can associate this operator with a descendant of a $\mc{B}^{1/2}_{1}$ supermultiplet, which we will refer to as the \textit{tilt supermultiplet}. The remaining states in this supermultiplet can be combined into a $\psi_{a\dot{a}}$ spinor. The structure of the multiplet is 
\begin{align}
    \mc{B}^{1/2}_{1}:\qquad {[\frac 12,1,1]}\rightarrow [1,0, 0] \oplus [1,0, 2] 
\end{align}
As for the supersymmetry transformations, we have
\begin{align}
    Q_{a\dot{a}} \psi_{b\dot{b}}&=  \epsilon_{ab} \varphi_{\dot{a}\dot{b}} \,, \nonumber\\
     Q_{a\dot{a}} \varphi_{\dot{b}\dot{c}}&= -\, \epsilon_{\dot{a}\dot{b}} \partial_t\psi_{a\dot{c}} \,.
\end{align}

Finally, let us notice that the displacement and tilt supermultiplets combine into one bigger supermultiplet in the more general $\mathfrak{d}(2,1;\sin^2{\Omega})$ algebra, which for $\Omega \rightarrow 0$ or $\Omega \rightarrow \frac{\pi}{2}$ reduces to the $\mathfrak{psu}(1,1|2)\times \mathfrak{su}(2)_A$ algebra of our defect. 

\subsection{Topological sector}
\label{sec: topological sector}

In this section we show the existence of a topological sector for correlators of primaries of the ${\cal B}^{\Delta}_{\ell}$ short multiplets. This will prove useful later when studying the OPE decomposition of such correlators, as it will allow us to derive the structure of the superconformal blocks associated to such OPE expansion. 

Let us begin the analysis by noticing that it is possible to define \textit{twisted} conformal generators $\hat{P},\hat{K}$ and $\hat{D}$ as
\begin{equation}
\hat{P}=P +  R^+_- \,, \qquad \hat{K}=K-  R^-_+ \,, \qquad \text{and} \qquad \hat{D}=D-R_0 \,.
\end{equation}

Crucially, these operators are $\mathbb{Q}$-exact, which can be seen from
\begin{align}
	\hat{P} &= -\{ \mathbb{Q}_{\dot{+}} ,Q_{-\dot{-}} \} =
 \{ \mathbb{Q}_{\dot{-}} ,  Q_{-\dot{+}} \} \,,
\nonumber \\
	\hat{K} &= \{ \mathbb{Q}_{\dot{+}} ,S_{+\dot{-}} \} =-\{ \mathbb{Q}_{\dot{-}} ,S_{+\dot{+}} \} \,, \nonumber \\
 \hat{D} &= -\frac{1}{2}\{ \mathbb{Q}_{\dot{+}} ,\mathbb{Q}_{\dot{+}}^{\dagger} \} =-\frac{1}{2} \{ \mathbb{Q}_{\dot{-}} ,\mathbb{Q}_{\dot{-}}^{\dagger} \} \,,
\end{align}
with
\begin{equation}
\mathbb{Q}_{\dot{a}}=Q_{+\dot{a}}+S_{-\dot{a}} \,,
\end{equation}
and where we are using that
\begin{equation}
Q_{a\dot{a}}^{\dagger}=S^{a\dot{a}} \,.
\end{equation}

Let us consider the superprimary operator ${\cal O}^{\Delta}_{\ell}(t,Y,W)$ of a ${\cal B}^{\Delta}_{\ell}$ short multiplet, where $t$ is the time coordinate, and $Y^{a}$ and $W^{\dot{a}}$ are auxiliary variables that are respectively introduced to contract every possible R-symmetry and automorphism index in the superprimary field. Then, we have that
\begin{equation}
\label{superprimary annihilation}
    [\mathbb{Q}_{\dot{a}}, {\cal O}^{\Delta}_{\ell} (0,0,W)]=0 \,.
\end{equation}
At this point it is useful to define
\begin{equation}
\label{timeev}
\hat{{\cal O}}^{\Delta}_{\ell} (t, W) := e^{-t \hat{P}} {\cal O}^{\Delta}_{\ell} (0,0,W)e^{t \hat{P}} \,.
\end{equation}
Given that $\hat{P}$ is $\mathbb{Q}$-exact, it commutes with $\mathbb{Q}_{\dot{a}}$. Therefore,
\begin{equation}
\label{superprimary annihilation 2}
    [\mathbb{Q}_{\dot{a}}, {\cal O}^{\Delta}_{\ell} (t,W)]=0 \,.
\end{equation}
Interestingly, when restricting to the frame\footnote{We can always do this given that correlators are homogeneous functions of the $Y$ variables \cite{Dolan:2004mu}. In this frame we get $Y_{ij}=y_i-y_j$.}
\begin{equation}
\label{top frame}
    Y(y)=(y,1) \,,
\end{equation}
we obtain
\begin{equation}
\hat{{\cal O}}^{\Delta}_{\ell} (t, W)= {\cal O}^{\Delta}_{\ell} (t,Y(t),W) \,.
\end{equation}
Then, we see that the $\hat{{\cal O}}^{\Delta}_{\ell}$ operators satisfy
\begin{align}
\partial_{t_1} \langle \hat{{\cal O}}^{\Delta}_{\ell} (t_1,W_1) \dots \hat{{\cal O}}^{\Delta}_{\ell} (t_n,W_n) \rangle &= -\langle [\hat{P}, \hat{{\cal O}}^{\Delta}_{\ell} (t_1,W_1)] \dots \hat{{\cal O}}^{\Delta}_{\ell} (t_n,W_n) \rangle  \nonumber \\
&=\langle [\{ \mathbb{Q}_{\dot{+}}, Q_{-\dot{-}} \}, \hat{{\cal O}}^{\Delta}_{\ell} (t_1,W_1)] \dots \hat{{\cal O}}^{\Delta}_{\ell} (t_n,W_n) \rangle \nonumber \\
&=0 \,,
\end{align}
where in the last step we have used \eqref{superprimary annihilation 2} and  $\mathbb{Q}_{\dot{+}} |0\rangle= \mathbb{Q}_{\dot{+}}^\dagger |0\rangle=0 $.
Therefore, we can conclude that the correlation functions of superprimaries of 1/2 BPS multiplets are topological when taking $Y_i(t_i)=(t_i,1)$, i.e.
\begin{align}
\label{top sector}
\partial_{t_i} \langle {\cal O}^{\Delta}_{\ell} (t_1,Y_1(t_1),W_1) \dots {\cal O}^{\Delta}_{\ell} (t_n, Y_n(t_n),W_n) \rangle &= 0 \,.
\end{align}

For a four-point function this expression will become particularly useful. Due to conformal and R-symmetry, the tensorial structure and time dependence of four-point correlators is known up to functions of the cross ratios $\chi$, $\rho_1$ and $\rho_2$  associated to the coordinates $t$, $Y$ and $W$, respectively, which we define as
\begin{equation}
\label{cross ratio defs}
    \chi= \frac{t_{12}t_{34}}{t_{13}t_{24}} \,, \qquad \rho_1= \frac{Y_{12}Y_{34}}{Y_{13}Y_{24}} \,, \qquad \rho_2= \frac{W_{12}W_{34}}{W_{13}W_{24}}\,,
\end{equation}
with $t_{ij}=t_i-t_j$, $Y_{ij}=\epsilon_{ab} Y^a Y^b$ and $W_{ij}=\epsilon_{\dot{a} \dot{b}} W^{\dot{a}} W^{\dot{b}}$. For example,
\begin{align}
\label{top4pt}
\langle {\cal O}^{\Delta}_{\ell} (t_1,Y_1,W_1) {\cal O}^{\Delta}_{\ell} (t_2,Y_2,W_2) {\cal O}^{\Delta}_{\ell} (t_3,Y_3,W_3) & {\cal O}^{\Delta}_{\ell} (t_4, Y_4,W_4) \rangle \nonumber \\
   =&\left(\frac{Y_{12}Y_{34}}{t_{12}t_{34}}\right)^\Delta (W_{12}W_{34})^{\ell} f(\chi,\rho_1,\rho_2) 
\end{align}
The form of the function $f(\chi,\rho_1,\rho_2)$ is constrained by the existence of a topological sector. This can be seen by evaluating the four-point function  \eqref{top4pt} in the frame \eqref{top frame}, as then the condition \eqref{top sector} implies
\begin{align}
\label{topfin}
  \partial_\chi f(\chi,\chi,\rho_2)=0 \,.
\end{align}

\section{Two- and three-point functions}
\label{sec 2-pt and 3-pt}

Correlation functions are strongly constrained by superconformal symmetry. For two- and three-point functions, conformal symmetry completely fixes the time dependence of correlators up to overall constants that depend on the operators included in the correlation function. On top of this, for correlators of fields belonging to the same supermultiplet, those constants are further related by supersymmetry. As supermultiplets contain degrees of freedom with different spin, it is easy to identify correlators that will be trivially zero. A simple example is the two-point function of a fermion and a scalar. Despite being trivial, by acting with supersymmetry transformations on such correlators one can obtain relations between the two-point functions of two scalars and those of two fermions (with some derivatives). This procedure can be generalized to different $n$-point functions. In this section we will apply it to obtain the two- and three-point functions for operators in the tilt and displacement supermultiplets.

\subsection{Tilt supermultiplet}

Let us begin with the tilt supermultiplet. The most general ansatz that one can in principle write for the two-point functions is 
\begin{align}
     \langle \psi_{a\dot{a}}(t_1) \psi_{b\dot{b}}(t_2)\rangle &=\frac{\epsilon_{ab}\epsilon_{\dot{a}\dot{b}}}{t_{12}}  \,, \nonumber \\
     \langle \varphi_{\dot{a}\dot{b}}(t_1)\varphi_{\dot{c} \dot{d}}(t_2) \rangle &=  \frac{\kappa_1 \, \epsilon_{\dot{a}\dot{c}}\epsilon_{\dot{b}\dot{d}}+\kappa_2 \,\epsilon_{\dot{a}\dot{d}}\epsilon_{\dot{b}\dot{c}}+\kappa_3 \,\epsilon_{\dot{a}\dot{b}}\epsilon_{\dot{c}\dot{d}}}{t_{12}^2} \,,
\end{align}
where we have arbitrarily chosen a unit normalization for the superprimary. We can fix the unknown constants in the above ansatz by noticing that
\begin{equation}
0= Q_{a \dot{a}} \langle \psi_{b\dot{b}}(t_1) \phi_{\dot{c}\dot{d}}(t_2)\rangle=\epsilon_{a b} \langle \varphi_{\dot{a}\dot{b}}(t_1) \varphi_{\dot{c}\dot{d}}(t_2)\rangle + \epsilon_{\dot{a}\dot{c}}  \langle \psi_{b\dot{b}}(t_1) \psi_{a\dot{d}}(t_2)\rangle \,,
\end{equation}
which gives us
\begin{align}
\label{tilt2pt}
     \langle \psi_{a\dot{a}}(t_1) \psi_{b\dot{b}}(t_2)\rangle &=\frac{\epsilon_{ab}\epsilon_{\dot{a}\dot{b}}}{t_{12}}  \,,  \nonumber \\
     \langle \varphi_{\dot{a}\dot{b}}(t_1)\varphi_{\dot{c} \dot{d}}(t_2) \rangle &=  \frac{\epsilon_{\dot{a}\dot{c}}\epsilon_{\dot{b}\dot{d}}}{t_{12}^2} \,.
\end{align}

Applying the same procedure to determine relations among the three-point functions of the tilt multiplet one obtains that all the three-point functions of the this supermultiplet are vanishing. This is the natural consequence of the vanishing of the three-point function
of the $\psi_{a\dot{a}}$ superprimary.
x|

\subsection{Displacement supermultiplet}

Let us now turn to the two- and three-point functions of the displacement multiplet. Imposing supersymmetry we can constrain the corresponding two-point functions to be 
\begin{align}
    \langle \phi_{ab}(t_1)\phi_{cd}(t_2) \rangle &=  -\frac{1}{2} \frac{\epsilon_{ac}\epsilon_{bd}+\epsilon_{ad}\epsilon_{bc}}{t_{12}^2}  \,,  \nonumber \\
     \langle \chi_{a\dot{a}}(t_1) \chi_{b\dot{b}}(t_2)\rangle &=2\frac{\epsilon_{ab}\epsilon_{\dot{a}\dot{b}}}{t_{12}^3}  \,,  \nonumber \\
     \langle \rho(t_1)\rho(t_2) \rangle &= \frac{3}{t_{12}^4}  \,,
     \label{rhonorm}
\end{align}
where again we have arbitrarily fixed the normalization of the superprimary. As for the 
three-point functions, we get
\begin{align}
    \langle \phi_{ab}(t_1)\phi_{cd}(t_2) \phi_{ef}(t_3) \rangle &=  \frac{\sigma \,(\epsilon_{ad}\epsilon_{cf}\epsilon_{eb}-\epsilon_{af}\epsilon_{cb}\epsilon_{ed})}{t_{12} t_{23} t_{13}} \,, \nonumber \\
     \langle \rho(t_1)\rho(t_2) \rho(t_3) \rangle &=\frac{6\, \sigma}{t_{12}^2t_{23}^2t_{13}^2} \,, \nonumber \\
     \langle \phi_{ab}(t_1)\phi_{cd}(t_2) \rho(t_3) \rangle &= \frac{-\frac{\sigma}{2} (\epsilon_{ac}\epsilon_{bd}+\epsilon_{bc}\epsilon_{ad})}{t_{23}^2t_{13}^2} \,, \nonumber \\
      \langle \chi_{a\dot{a}}(t_1) \chi_{b\dot{b}}(t_2) \phi_{cd}(t_3) \rangle &= -\frac{\sigma \, (\epsilon_{ac} \epsilon_{bd}\epsilon_{\dot{a}\dot{b}}+\epsilon_{ad} \epsilon_{bc}\epsilon_{\dot{a}\dot{b}})}{t_{12}^2t_{23}t_{13}} \,, \nonumber \\
        \langle \chi_{a\dot{a}}(t_1) \chi_{b\dot{b}}(t_2) \rho(t_3) \rangle &= \frac{3 \sigma \, \epsilon_{ab}\epsilon_{\dot{a}\dot{b}}}{t_{12}t_{23}^2t_{13}^2} \,,
        \label{3ptdispl}
\end{align}
where $\sigma$ is a constant that cannot be fixed with superconformal symmetry.

\section{Four-point functions}
\label{sec 4-pt}

In this section we will turn to the computation of the four-point functions of fields within the displacement and tilt supermultiplets. As it is well known, the time dependence of four-point functions is constrained by conformal invariance up to functions of a conformal cross-ratio. However, supersymmetry imposes non-trivial constraints between the cross-ratio dependencies of correlators within a same supermultiplet. In particular, we will show that we can use supersymmetry to express every correlator of insertions within the displacement supermultiplet in terms of two unknown cross-ratio functions and a constant. Similarly, correlators of the tilt supermultiplet will be completely determined by two cross-ratio functions and two constants. In order to compute those functions and constants we will perform an analytic bootstrap analysis, for which we will impose several consistency conditions on the correlators. We obtain that, up to next-to-leading order in the strong coupling expansion, the four-point functions of the displacement and tilt supermultiplets can be completely fixed in terms of two parameters, for which we will provide an holographic interpretation in section \ref{Sec: Holographic description}.

\subsection{Constraints from supersymmetry}

Let us begin by studying the constraints that supersymmetry imposes within four-point functions of fields that belong to a same supermultiplet. In particular, we will begin our analysis by focusing on the case of the displacement supermultiplet.

To write down all possible all the functions of the cross-ratio that enter in the four-point correlators of components of the displacement multiplet, it is useful to define the following tensor structures 
\begin{align}
S^{(2)}_{ab,cd} &\ := \epsilon_{ac}\epsilon_{bd}+\epsilon_{ad}\epsilon_{bc}\,,
\nonumber\\
S^{(3)}_{ab,cd,ef} &\ := \epsilon_{ad}\epsilon_{cf}\epsilon_{eb}+\epsilon_{af}\epsilon_{bc}\epsilon_{ed}\,.
\end{align}
In terms of them, we get that R-symmetry implies, for example
\begin{align}
\langle \phi_{a_1b_1}(t_1)\phi_{a_2b_2}(t_2)\phi_{a_3b_3}(t_3)\phi_{a_4b_4}(t_4)\rangle
= & \ \frac{1}{t_{12}^2t_{34}^2}\left( 
S^{(2)}_{a_1b_1,a_2b_2}\, S^{(2)}_{a_3b_3,a_4b_4} \; d_1(\chi)\right.
\nonumber\\
& \hspace{-4cm}
\left. + S^{(2)}_{a_1b_1,a_3b_3}\, S^{(2)}_{a_2b_2,a_4b_4} \; d_2(\chi)
 +       S^{(2)}_{a_1b_1,a_4b_4}\, S^{(2)}_{a_2b_2,a_3b_3} \; d_3(\chi)\right) \,,
 \label{phi4}
\\
\langle \phi_{a_1b_1}(t_1)\phi_{a_2b_2}(t_2)\chi_{a_3\dot a_3}(t_3)\chi_{a_4\dot a_4}(x_4)\rangle
= & \ \frac{\epsilon_{\dot a_3\dot a_4}}{t_{12}^2t_{34}^3}\left(S^{(2)}_{a_1b_1,a_2b_2}\,\epsilon_{a_3 a_4} \; d_4(\chi)
\right.
\nonumber\\
& \left. 
\hspace{-4cm} +(S^{(2)}_{a_1b_1,a_2a_3}\,\epsilon_{b_2 a_4} + S^{(2)}_{a_1b_1,b_2a_3}\,\epsilon_{a_2 a_4} ) \; d_5(\chi) \right) \,,
\nonumber\label{phi2chi2}
\\
\nonumber \langle \phi_{a_1b_1}(t_1)\phi_{a_2b_2}(t_2)\phi_{a_3b_3}(t_3)\rho(t_4)\rangle
= & \ \frac{1}{t_{12}t_{14}t_{24}t_{34}^2}
S^{(3)}_{a_1b_2,a_2b_2,a_3b_3} d_6(\chi) \,,
\end{align}
where $\chi$ is one of the cross ratios defined in \eqref{cross ratio defs}. Similarly, we can use R-symmetry to constrain all the remaining correlators of components of the displacement supermultiplet, getting a total of 14 functions of the cross-ratio $\chi$. As we have done for 2-point and 3-point functions, we can find relations between these functions by considering the supersymmetry transformations of trivial four-point functions. For example, by studying the behaviour under supersymmetry  of $\langle \phi_{a_1b_1}(t_1)\phi_{a_2b_2}(t_2)\phi_{a_3 b_3}(t_3)\chi_{a_4\dot a_4}(x_4)\rangle$ we get
\begin{equation}
\left(d_1(\chi)+ \frac{1}{\chi^2} d_2(\chi)+  \frac{(1 - \chi)^2}{\chi^2} d_3(\chi) 
\right)'=0 \,,
\label{susyconditiondisp}
\end{equation}
which allows us to easily express $d_3(\chi)$ in terms of $d_1(\chi)$, $d_2(\chi)$ and an integration constant $D_0$
\begin{equation}
d_3(\chi)=\frac{D_0 \chi^2 - 4 \chi^2 d_1(\chi) - 4 d_2(\chi)}{4 (1 - \chi)^2} \,.
\end{equation}
Moreover, we obtain that
\begin{align}
d_4(\chi)= &\ 
-4 d_1(\chi) + \frac{4 d_2(\chi)}{1 - \chi} + \frac{(2 - 4 \chi) \chi d_1'(\chi)}{1 - \chi} - \frac{2 \chi d_2'(\chi)}{1 - \chi} \,,
\nonumber\\
d_5(\chi)= &\
-\frac{2 (2 - \chi) d_2(\chi)}{1 - \chi} + \frac{\chi^2 d_1'(\chi)}{1 - \chi} + \frac{(2 - \chi) \chi d_2'(\chi)}{1 - \chi} \,,
\nonumber\\
d_6(\chi)= &\
-\frac{4 d_2(\chi)}{\chi} + 2\chi d_1'(\chi) + 2d_2'(\chi)\,.
\end{align}
Proceeding in the same way with other trivial 4-point correlators we can fix the remaining functions of the cross-ratio in terms of $d_1(\chi)$, $d_2(\chi)$ and the integration constant $D_0$. In particular, for correlator of 4 displacement operators we get
\begin{equation}
\langle \rho(t_1)\rho(t_2)\rho(t_3)\rho(t_4)
\rangle =     
\frac{d_{14}(\chi)}{t_{12}^4t_{34}^4} \,,
\label{rho4pt}
\end{equation}
with
\begin{align}
d_{14}(\chi) & =
\frac{9 D_0 \chi^4}{(-1 + \chi)^4} 
- \frac{36 (-1 + 4 \chi - 6 \chi^2 + 4 \chi^3) d_1(\chi)}{(-1 + \chi)^4} 
+ \frac{4 \chi (-14 + 43 \chi - 46 \chi^2 + 8 \chi^3) d_2(\chi)}{(-1 + \chi)^4} 
\nonumber
\\
& - \frac{2 \chi (16 - 65 \chi + 86 \chi^2 - 13 \chi^3 - 36 \chi^4 + 12 \chi^5) d_1'(\chi)}{(-1 + \chi)^4} 
\nonumber
\\
&
 + \frac{2 \chi^2 (16 - 50 \chi + 33 \chi^2 + 6 \chi^3 - 7 \chi^4 + 2 \chi^5) d_2'(\chi)}{(-1 + \chi)^4} 
- \frac{2 \chi^2 (-7 + 15 \chi - 3 \chi^2 + 2 \chi^3) d_1''(\chi)}{(-1 + \chi)^2} 
\nonumber
\\
&
+ \frac{2 \chi^3 (-2 + 3 \chi - 15 \chi^2 + 7 \chi^3) d_2''(\chi)}{(-1 + \chi)^2} 
+ \frac{4 \chi^3 (1 - \chi + \chi^2) d_1'''(\chi)}{(-1 + \chi)} 
+ \frac{4 \chi^4 (3 - 4 \chi + 2 \chi^2) d_2'''(\chi)}{(-1 + \chi)} 
\nonumber
\\
&
- \chi^4 (-1 + 2 \chi) d_1''''(\chi) 
+ \frac{(-2 \chi^5 + 9 \chi^6 - 16 \chi^7 + 14 \chi^8 - 6 \chi^9 + \chi^{10}) d_2''''(\chi)}{(-1 + \chi)^4} \,.
\label{d14}
\end{align}

In the following sections we will perform a bootstrap analysis that will allow us to compute the unknown functions $d_1(\chi)$ and $d_2(\chi)$. In order to do so, we will find useful to work with correlators that are invariant under R-symmetry and the automorphism. As in Section \ref{sec: topological sector}, we can easily do this by introducing two auxiliary variables $Y^a$ and $W^{\dot{a}}$ in the fundamental representation of the R-symmetry and automorphism groups, respectively. For example, by defining
\begin{equation}
\Phi(t,Y):= Y^a Y^b \phi_{ab}(t) \,,
\end{equation}
we obtain
\begin{align}
\label{4-pt displ primary}
\langle \Phi(t_1,Y_1) \Phi(t_2,Y_2) \Phi(t_3,Y_3) \Phi(t_4,Y_4) \rangle &= \frac{Y_{12}^2 Y_{34}^2}{t_{12}^2 t_{34}^2} \, \mathbf{f}(\chi,\rho_1) \,,
\end{align}
where $\rho_1$ was defined in \eqref{cross ratio defs} and with
\begin{equation}
\label{4-pt displ Ward}
\mathbf{f}(\chi,\rho_1)= D_0 + \left( 1 - \frac{\chi}{\rho_1}\right) f_1(\chi) + \left( 1 - \frac{\chi}{\rho_1}\right)^2 f_2(\chi) \,,
\end{equation}
where
\begin{align}
f_1(\chi)&= \frac{2 D_0}{\chi -1}-\frac{8 \, d_1(\chi )}{\chi -1}-\frac{8 \, d_2(\chi )}{(\chi -1) \chi } \,,
\nonumber
\\
f_2(\chi)&= \frac{D_0}{(\chi -1)^2}-\frac{4 d_1(\chi )}{(\chi -1)^2}+ \frac{d_2(\chi)}{(\chi -1)^2} \left( 4-\frac{8}{\chi }\right) \,.
\label{dstof2}
\end{align}

Finally, let us turn to the constraints imposed by supersymmetry on the correlators of the tilt supermultiplet. Proceeding in a similar way as we have done for the displacement supermultiplet, we obtain that all the correlators of the tilt supermultiplet can be expressed in terms of two constants $T_1$ and $T_2$ and two functions $h_1(\chi)$ and $h_2(\chi)$ of the cross-ratio $\chi$. In particular, for the superprimary $\psi_{a \dot{a}}$ we get
\begin{align}
\label{4 pt tilt primary}
\langle \Psi(t_1,Y_1,W_1) \Psi(t_2,Y_2,W_2) &\Psi(t_3,Y_3,W_3) \Psi(t_4,Y_4,W_4) \rangle = \frac{Y_{12} Y_{34} W_{12}W_{34}}{t_{12} t_{34}} \, \mathbf{h}(\chi,\rho_1,\rho_2) \,,
\end{align}
where $\rho_2$ is one of the cross ratios introduced in \eqref{cross ratio defs}, and with
\begin{equation}
\label{4-pt tilt Ward}
\mathbf{h}(\chi,\rho_1,\rho_2)= T_1 + \left( 1 - \frac{\chi}{\rho_1}\right) h_1(\chi) + \frac{1}{\rho_2} \left(T_2+ \left( 1 - \frac{\chi}{\rho_1}\right) h_2(\chi) \right) \,,
\end{equation}
and where
\begin{equation}
\Psi(t,Y,W):= Y^a W^{\dot{a}} \psi_{a\dot{a}}(t) \,.
\end{equation}

It is worth noticing that the condition \eqref{topfin}, which was derived from the existence of a topological sector, it is verified straightforwardly by the expressions \eqref{4-pt displ Ward} and \eqref{4-pt tilt Ward}, derived from supersymmetry constraints. We would also like to point out that the structure imposed by supersymmetry in \eqref{4-pt displ Ward} and \eqref{4-pt tilt Ward} can also be found from the constraints of appropriate superconformal Ward identities \cite{Dolan:2004mu}.

To compare with holographic results we will be interested in the scalar four-point function
\begin{equation}
\langle \hat{\varphi} (t_1) \hat{\varphi} (t_2) \hat{\varphi} (t_3) \hat{\varphi} (t_4) \rangle= \frac{h_3 (\chi)}{t_{12}^2 t_{34}^2} \,,
\label{varphi4pt}
\end{equation}
defined by the contraction
\begin{equation}
\hat{\varphi}(t)= \epsilon^{\dot{a}\dot{b}} \varphi_{\dot{a}\dot{b}} (t) \,.
\label{varphidef}
\end{equation}
Supersymmetry relates $h_3$ with the functions appearing in the super primary through 
\begin{align}
    h_3(\chi)&= \left(2 \chi ^4-6 \chi ^3+4 \chi ^2\right) h_1''(\chi )+\left(6 \chi ^3-4 \chi ^2-4 \chi \right) h_1'(\chi )+\left(2 \chi ^2+4\right) h_1(\chi
   ) \nonumber\\
   &\quad +\left(4 \chi ^4-6 \chi ^3+2 \chi ^2\right) h_2''(\chi )+\left(12 \chi ^3-8 \chi ^2-2 \chi \right) h_2'(\chi )+\left(4 \chi ^2+2\right)
   h_2(\chi ) \nonumber \\
   &\quad +2 T_1+4 T_2 \,.
   \label{h3}
   \end{align}

\subsection{Operator product expansion and superconformal blocks}

In the following we will perform a bootstrap computation of the \eqref{4-pt displ Ward} and \eqref{4-pt tilt Ward} four-point functions. In order to do so, it will prove crucial to expand those correlators in terms of exchanged operators using the operator product expansion (OPE). 

In particular, for the bootstrap analysis of \eqref{4-pt displ Ward} and \eqref{4-pt tilt Ward} we will be interested in the $\mathcal{B}^{1/2}_{1} \times \mathcal{B}^{1/2}_{1}$ and $\mathcal{B}^{1}_{0} \times \mathcal{B}^{1}_{0}$ OPEs. Using the selection rules for the R-symmetry and automorphism $SU(2)$'s along with the operator content of the theory described in Section \ref{reps} we get\footnote{The long operators satisfy the unitarity bound $\Delta > \frac{r}{2}$, whereas the $\Delta=\frac r 2$ bound is saturated by the short 1/2 BPS supermultiplets.}
 \begin{align}
 \nonumber
     &\mathcal{B}^{1/2}_{1} \times \mathcal{B}^{1/2}_{1} \subset \mc{I}+\mathcal{B}^{1}_{0}+\mathcal{B}^{1}_{2}+\sum_{\Delta>0} \left(  \mathcal{L}_{0,0}^\Delta+\mathcal{L}_{0,2}^\Delta+\mathcal{L}_{2,2}^\Delta+\mathcal{L}_{2,0}^\Delta \right) \,, \\
     \label{displ ope}
     &\mathcal{B}^{1}_{0} \times \mathcal{B}^{1}_{0} \subset \mc{I}+\mathcal{B}^{1}_{0}  +\mathcal{B}^{2}_{0} + \sum_{\Delta>0} \left(  \mathcal{L}_{0,0}^\Delta+\mathcal{L}_{2,0}^\Delta+\mathcal{L}_{4,0}^\Delta \right) \,.
\end{align}
Translating the above OPEs into superconformal block expansions for the $\mathbf{h}$ and $\mathbf{f}$ functions defined in \eqref{4-pt displ Ward} and \eqref{4-pt tilt Ward} we obtain
\begin{align}
\nonumber
\mathbf{h}&= b_{\cal I}+ b_{1,0} \, \mathcal{G}_{\mc{B}^{1}_{0}}+  b_{1,2} \, \mathcal{G}_{\mc{B}^{1}_{2}}+ \sum_{\Delta>0} \left(  c_{\Delta,0,0} \, \mathcal{G}_{\mc{L}^{\Delta}_{0,0}}+ c_{\Delta,0,2} \, \mathcal{G}_{\mc{L}^{\Delta}_{0,2}}+ c_{\Delta,2,2} \, \mathcal{G}_{\mc{L}^{\Delta}_{2,2}}+ c_{\Delta,2,0} \, \mathcal{G}_{\mc{L}^{\Delta}_{2,0}}  \right) \,, \\
\label{f ope 1}
\mathbf{f}&=  \tilde{b}_{\cal I}+ \tilde{b}_{1,0} \, \mathcal{G}_{\mc{B}^{1}_{0}} + \tilde{b}_{2,0}  \, \mathcal{G}_{\mc{B}^{2}_{0}}+ \sum_{\Delta>0} \left( \tilde{c}_{\Delta,0,0} \,  \mathcal{G}_{\mc{L}^{\Delta}_{0,0}} + \tilde{c}_{\Delta, 2,0} \, \mathcal{G}_{\mc{L}^{\Delta}_{2,0}} + \tilde{c}_{\Delta,4,0} \, \mathcal{G}_{\mc{L}^{\Delta}_{4,0}} \right) \,,
\end{align}
where $\mathcal{G}_{\mc{B}^{\Delta}_{\ell}}$ and $\mathcal{G}_{\mc{L}^{\Delta}_{r,\ell}}$ are the superconformal blocks associated to short and long supermultiplets, respectively, and $b_{\Delta, \ell}, \, \tilde{b}_{\Delta, \ell}, \, c_{\Delta, r, \ell}, \, \tilde{c}_{\Delta,r, \ell}$ are their corresponding OPE coefficients\footnote{Recall the shortening condition $r=2\Delta$ for the short multiplets.}. 
Generically, we can expand the superblocks as
\begin{align}
\label{generic exp 1}
\mathcal{G}_{\mc{B}^{\Delta}_{\ell}} 
   & = \sum_i \alpha^{\Delta,i}_{\ell} G^{\Delta_i}_{r_i,\ell_i}(\rho_1,\rho_2,\chi) \,,
   \\
   \label{generic exp 2}
   \mathcal{G}_{\mc{L}^{\Delta}_{r,\ell}}
   & = \sum_i \beta^{\Delta,i}_{r,\ell} G^{\Delta_i}_{r_i,\ell_i}(\rho_1,\rho_2,\chi) \,,
\end{align}
for some set of coefficients $\alpha^{\Delta,i}_{\ell}$ and $\beta^{\Delta,i}_{r,\ell}$ and where the sum runs over all conformal primaries in the respective supermultiplet, with
\begin{equation}
\label{conf block def}
G^\Delta_{r,\ell} = g_{-\frac r 2}(\rho_1)g_{-\frac \ell 2}(\rho_2) g_\Delta(\chi) \,, \quad g_{k}(x)=x^{k} \, _2F_1(k ,k ;2 k ;x) \,.
\end{equation}
In order to constrain \eqref{generic exp 1} and \eqref{generic exp 2} we should first note that odd-level descendants should give a vanishing contribution, since they correspond to three-point functions with an odd number of fermions. For example, for the $\mc{B}^{\Delta}_{\ell}$ short supermultiplets we only need to consider the contributions from the superprimary and the level-two superdescendant, i.e. from the red multiplets in 
\begin{align}
    \mc{B}^{\Delta}_{\ell}:\qquad {\color{red}{[\Delta,2\Delta,\ell]}}\rightarrow [\Delta+\frac 12,2\Delta-1,\ell\pm 1]\rightarrow {\color{red}{[\Delta+ 1,2\Delta-2,\ell]}}
\end{align}
Similarly, for long operators we have the contributions of the superprimary and the superdescendants at level two and four, i.e.
\begin{align}
   \mc{L}^\Delta_{r,\ell}:\hspace{5cm}\color{red}[\Delta,&\color{red}r,\ell]\nonumber \\
    [\Delta+\frac 12,r+1,\ell+1] \qquad [\Delta+\frac 12,r+1,\ell-1] &\qquad [\Delta+\frac 12,r-1,\ell+1] \qquad [\Delta+\frac 12,r-1,\ell-1] \nonumber\\
    \color{red}[\Delta+1,r\pm2,\ell]\qquad 2 \, [\Delta+1,&\color{red}r,\ell]\qquad[\Delta+1,r,\ell\pm2]\\
   [\Delta+\frac 32,r+1,\ell+1] \qquad [\Delta+\frac 32,r+1,\ell-1] &\qquad [\Delta+\frac 32,r-1,\ell+1] \qquad [\Delta+\frac 32,r-1,\ell-1] \nonumber\\
    \color{red}[\Delta+2,&\color{red}r,\ell] \nonumber
\end{align}

At this point it proves useful to impose the existence of a topological sector for correlators of superprimaries of the $\mc{B}^{\Delta}_{\ell}$ supermultiplets, as discussed in Section \ref{sec: topological sector}, by requiring 
\begin{align}
\nonumber 
    \partial_{\chi} \mathbf{h} |_{\rho_1=\chi}&=0 \,, \\
\label{top sector f}
    \partial_{\chi} \mathbf{f} |_{\rho_1=\chi}&=0 \,.
\end{align}
Equivalently, one can use the superconformal Ward identities presented in \cite{Baume:2019aid, Bliard:2024und} or the constraints \eqref{susyconditiondisp} derived above. Let us note that \eqref{top sector f}   
are symmetry constraints that do not depend on the dynamics of the theory. Therefore, imposing \eqref{top sector f} on \eqref{f ope 1} we see that each superconformal block contributing to the \eqref{f ope 1} expansions should independently satisfy the topological sector constraints, regardless of the values of the OPE coefficients. Furthermore, those constraints should be valid for four-point functions of arbitrary $\mc{B}^{\Delta}_{\ell}$ superprimaries, and therefore we obtain
\begin{align}
\nonumber 
\mc{G}_{\mc{B}^{\Delta}_{\ell}}&=G^{\Delta}{}_{2\Delta,\ell}-\frac{\Delta^2}{4(4\Delta^2-1)} G^{\Delta+1}{}_{2\Delta-2,\ell} \,, \\
\label{L blocks}
\mc{G}_{\mc{L}^\Delta_{r,\ell}}&=G^{\Delta }{}_{r,\ell}-G^{\Delta +1}{}_{r+2,\ell}-\frac{r^2}{16(r^2-1)} G^{\Delta +1}{}_{r-2,\ell}+\frac{(\Delta +1)^2 G^{\Delta +2}{}_{r,\ell}}{4 (2 \Delta +1) (2 \Delta +3)} \,,
\end{align}
for generic $\mc{B}^{\Delta}_{\ell}$ and $\mc{L}^\Delta_{r,\ell}$ supermultiplets. Since the $SU(2)_A$ automorphism acts as a spectator in these superconformal blocks, we recover the structure derived in \cite{Baume:2019aid}.

Let us return now to \eqref{f ope 1}. From 
\eqref{L blocks} we see that we can use the $SU(2)_R$ and $SU(2)_A$ selection rules to further constrain the $\mathbf{h}$ and $\mathbf{f}$ OPEs. To be more specific, we note from \eqref{L blocks} that each $\mc{L}^\Delta_{r,\ell}$ multiplet gives a non-vanishing contribution from a conformal primary of spin $r+2$. Therefore, the $\mc{L}^\Delta_{2,2}$ and $\mc{L}^\Delta_{2,0}$ supermultiplets can not be part of the \eqref{f ope 1} expansion, as they would contribute with a spin 4 operator, in contradiction with the $SU(2)$ selection rules for the $\mathcal{B}^{1/2}_{1} \times \mathcal{B}^{1/2}_{1}$ OPE. Similarly, we can rule out the $\mc{L}^\Delta_{4,0}$ from the $\mathcal{B}^{1}_{0} \times \mathcal{B}^{1}_{0}$ OPE. Therefore, we arrive at\footnote{A similar derivation of the selection rules can be performed by an analysis of three point functions between two short and one long supermultiplets, see for example \cite{Eden:2001ec,Arutyunov:2001qw,Eden:2001wg,Ferrara:2001uj}.}
 \begin{align}
\nonumber 
&\mathcal{B}^{1/2}_{1} \times \mathcal{B}^{1/2}_{1}  \sim \mc{I}+\mathcal{B}^{1}_{0}+\mathcal{B}^{1}_{2}+ \sum_{\Delta>0} \left(  \mathcal{L}_{0,0}^\Delta+\mathcal{L}_{0,2}^\Delta \right) \,, \\
\label{B1 OPE}
     &\mathcal{B}^{1}_{0} \times \mathcal{B}^{1}_{0} \sim \mc{I}+\mathcal{B}^{1}_{0}  +\mathcal{B}^{2}_{0} + \sum_{\Delta>0} \left(  \mathcal{L}_{0,0}^\Delta+\mathcal{L}_{2,0}^\Delta \right) \,,
\end{align}
and
\begin{align}
\nonumber 
\mathbf{h}&= b_{\cal I}+ b_{1,0} \, \mathcal{G}_{\mc{B}^{1}_{0}}+ b_{1,2} \, \mathcal{G}_{\mc{B}^{1}_{2}}+ \sum_{\Delta>0} \left(  c_{\Delta,0,0} \, \mathcal{G}_{\mc{L}^{\Delta}_{0,0}}+ c_{\Delta, 0,2} \, \mathcal{G}_{\mc{L}^{\Delta}_{0,2}} \right) \,, \\
\label{f ope 2}
\mathbf{f}&=  \tilde{b}_{\cal I}+ \tilde{b}_{1,0} \, \mathcal{G}_{\mc{B}^{1}_{0}} + \tilde{b}_{2,0} \, \mathcal{G}_{\mc{B}^{2}_{0}}+ \sum_{\Delta>0} \left(  \tilde{c}_{\Delta,0,0} \, \mathcal{G}_{\mc{L}^{\Delta}_{0,0}} + \tilde{c}_{\Delta,2,0} \, \mathcal{G}_{\mc{L}^{\Delta}_{2,0}} \right) \,.
\end{align}

\subsection{Analytic bootstrap}

In the spirit of \cite{Liendo:2018ukf,Ferrero:2021bsb,Ferrero:2023znz,Ferrero:2023gnu,Bianchi:2020hsz,Pozzi:2024xnu}, in this section we will follow an \textit{analytic bootstrap} approach for the computation of \eqref{4-pt displ primary} and \eqref{4 pt tilt primary} at strong coupling. This method consists of deriving the correlators by imposing superconformal symmetry and other consistency conditions. For reviews applied to the cases of one-dimensional superconformal defects in $\mc{N}=4$ SYM and ABJM see \cite{Ferrero:2023gnu,Bliard:2023zpe}.

To be more specific, we will focus on the computation of the ${\bf f}(\chi,\rho_1)$ and ${\bf h}(\chi,\rho_1,\rho_2)$ functions that give the cross-ratio dependence of \eqref{4-pt displ primary} and \eqref{4 pt tilt primary}. We will focus in their strong coupling limit, and therefore we will study the expansions
\begin{align}
{\bf h}(\chi,\rho_1,\rho_2) &={\bf h}^{(0)}(\chi,\rho_1,\rho_2)+ \epsilon \, {\bf h}^{(1)}(\chi,\rho_1,\rho_2)+ \epsilon^2 \,  {\bf h}^{(2)}(\chi,\rho_1,\rho_2) +\dots,\nonumber\\
{\bf f}(\chi,\rho_1) &={\bf f}^{(0)}(\chi,\rho_1)+ \epsilon\, {\bf f}^{(1)}(\chi,\rho_1)+ \epsilon^2 \,  {\bf f}^{(2)}(\chi,\rho_1) +\dots \,,
\end{align}
where $\epsilon$ is a coupling constant proportional to the inverse of the string tension. When comparing with the OPE expansions given in \eqref{B1 OPE}, we get 
\begin{align}
b_{\Delta,\ell} &= b_{\Delta,\ell}^{(0)} + \epsilon \, b_{\Delta,\ell}^{(1)} + \dots \,, \nonumber \\
\tilde{b}_{\Delta,\ell} &= \tilde{b}_{\Delta,\ell}^{(0)} + \epsilon \, \tilde{b}_{\Delta,\ell}^{(1)} + \dots \,, \nonumber\\
c_{\Delta,r,\ell} &= c_{\Delta,r,\ell}^{(0)} + \epsilon \, c_{\Delta,r,\ell}^{(1)}+ \dots \,, \nonumber\\
\tilde{c}_{\Delta,r,\ell} &= \tilde{c}_{\Delta,r,\ell}^{(0)} + \epsilon \, \tilde{c}_{\Delta,r,\ell}^{(1)}  + \dots \,,
\end{align}
for the OPE coefficients. For the identity we will consider
$b_{\cal I}=\tilde{b}_{\cal I}=1$, which amounts to
arbitrarily setting a unit two-point function coefficient of the superprimaries at all orders in $\epsilon$. In particular, this 
is convenient to 
relate the OPE coefficients to the square of the constants appearing in three-point functions\footnote{For the holographic analysis this normalization is also convenient, ensuring that only connected Witten diagrams contribute.}. Similarly,
\begin{align}
\Delta_{r,\ell}(\epsilon) &= \Delta_{r,\ell}^{(0)} + \epsilon \, \gamma_{\Delta,r,\ell}^{(1)}+ \dots \nonumber \,, \\
\tilde{\Delta}_{r,\ell}(\epsilon) &= \tilde{\Delta}_{r,\ell}^{(0)} + \epsilon \, \tilde{\gamma}_{\Delta,r,\ell}^{(1)} + \dots \,,
\end{align}
where $\Delta_{r,\ell}(\epsilon)$ and $\tilde{\Delta}_{r,\ell}(\epsilon)$ are the scaling dimensions of the ${\cal L}^{\Delta}_{r,\ell}$ operators that contribute to \eqref{B1 OPE}. 

As discussed in previous sections, one can use supersymmetry to fully constrain the R-symmetry dependence of ${\bf f}$ and ${\bf h}$. In the end we are left with three constants $D_0,\, T_1$ and $T_2$ and four functions $f_1(\chi), f_2(\chi), h_1(\chi)$ and $h_2(\chi)$ that completely determine ${\bf f}$ and ${\bf h}$. 
All of them will be expanded in powers of the coupling constant $\epsilon$ as
\begin{align}
\nonumber
T_i &=T_i^{(0)} + \epsilon \, T_i^{(1)} + \dots \,, \\
\nonumber
D_0 &= D_0^{(0)} + \epsilon \, D_0^{(1)} + \dots \,, \\
\nonumber
h_i(\chi)&=h_i^{(0)}(\chi)+ \epsilon\,  h_i^{(1)}(\chi) +\dots \,, \\
\label{strong coupling fi}
f_i(\chi)&=f_i^{(0)}(\chi)+ \epsilon\,  f_i^{(1)}(\chi) +\dots,
\end{align}
for $i=1,2$.

\subsubsection{Leading order}
\label{sec LO}

From AdS/CFT we know that the strong coupling expansion of the \eqref{4-pt displ primary} and \eqref{4 pt tilt primary} correlators can be recast as an expansion in terms of Witten diagrams. Therefore, from Wick contractions we get
\begin{align}
\nonumber 
\mathbf{h}^{(0)}(\chi,\rho_1,\rho_2) &=  
1-\frac{\chi}{\rho_1\rho_2}+\frac{\chi(1-\rho_1)(1-\rho_2)}{(1-\chi)\rho_1\rho_2} \,, \\
\label{LO f}
\mathbf{f}^{(0)} (\chi,\rho_1)&=1+\frac{\chi ^2}{\rho_1^2}+\frac{(1-\rho_1)^2 \chi ^2}{\rho_1^2 (\chi -1)^2} \,,
\end{align}
at leading order. This fixes
\begin{align}
\nonumber
 D_0^{(0)} &= 3\,, &T_1^{(0)}& = 0\,, &T_2^{(0)}&=0 \,,
\\
\nonumber
h_1^{(0)} (\chi)  &= \frac{1}{1-\chi}\,, 
&h_2^{(0)} (\chi)&= -\frac{\chi}{1-\chi}  \,,& &\\
\label{LO fs}
f_1^{(0)}(\chi) &= -\frac{2(\chi-2)}{\chi-1}
\,, &f_2^{(0)}(\chi) &= 1 + \frac{1}{(\chi-1)^2} \,. & &
\end{align}
Then, from the above results we can read that the leading-order CFT data is
\begin{align}
b_{1,0}^{(0)}&=-1 ,  & & & b_{1,2}^{(0)}&=0, & & \nonumber\\
\tilde{b}_{1,0}^{(0)}&=0, & & & \tilde{b}_{2,0}^{(0)}&=2, & &\nonumber  \\
c_{\Delta,0,0}^{(0)}&= 
\frac{\Gamma (\Delta) \Gamma (\Delta +1)}{2\Gamma \left(2\Delta\right)}
&\text{for even } \Delta,&    & c_{\Delta,0,0}^{(0)} &= 0 &\text{for odd } \Delta,& \nonumber
\\
c_{\Delta,0,2}^{(0)}&=
-\frac{\Gamma (\Delta) \Gamma (\Delta +1)}{\Gamma \left(2\Delta\right)} 
&\text{for odd } \Delta,&
 &c_{\Delta,0,2}^{(0)}& = 0 &\text{for even } \Delta,&\nonumber\\
\tilde{c}_{\Delta,0,0}^{(0)}&=
\frac{(\Delta -1) \Delta  (\Delta +2) \Gamma (\Delta )^2}{2 \Gamma (2 \Delta )} 
&\text{for even } \Delta ,&   &\tilde{c}_{\Delta,0,0}^{(0)} &= 0 &\text{for odd } \Delta,& \nonumber
\\
\tilde{c}_{\Delta,2,0}^{(0)}&=-\frac{(\Delta +1) 
\Gamma (\Delta +1)^2}{\Gamma (2 \Delta )}
&\text{for odd } \Delta\geq 3,&
&\tilde{c}_{\Delta,2,0}^{(0)} &= 0 &\text{for even } \Delta,& \nonumber\\
\label{LO CFT data-7}
\tilde{c}_{1,2,0}^{(0)}&=0. & & & & & & 
\end{align}

\subsubsection{Next-to-leading order}

For the next-to-leading order (NLO) in the strong coupling expansion, the starting point of the analytic bootstrap process consists in writing a suitable ansatz for the correlators \cite{Ferrero:2019luz}. In our case, we will propose an ansatz for the $f_i^{(1)}(\chi)$ and $h_i^{(1)}(\chi)$ functions that were introduced in 
\eqref{strong coupling fi}. When comparing to higher-dimensional setups, here is where the first simplification occurs. Crucially, in one-dimensional CFTs there is only one independent conformal cross ratio that can be constructed out of the four time coordinates of the insertions, as opposed to the two independent cross ratios that one has to consider in higher dimensions.

In order to formulate an ansatz for $f_i(\chi)$ and $h_i(\chi)$ it is convenient to use the intuition coming from the AdS/CFT correspondence. As presented in \eqref{LO f}, at leading order $f_i^{(0)}(\chi)$ and $h_i^{(0)}(\chi)$ are rational functions, given that there is no vertex integral involved in their computations. When moving to subleading corrections we expect transcendental functions to appear in the ansatz. Borrowing intuition from the analysis of 1/2 BPS line defects in the ${\cal N}=4$ super Yang-Mills and ABJM theories \cite{Liendo:2018ukf,Ferrero:2021bsb,Ferrero:2023znz,Ferrero:2023gnu,Bianchi:2020hsz,Pozzi:2024xnu}, we expect the N$^{L}$LO correction to $f_i(\chi)$ and $h_i(\chi)$ to be expressed in terms of harmonic polylogarithms \cite{Remiddi:1999ew, Alday:2015eya, Ferrero:2019luz} of transcendental weight less or equal to $L$\footnote{This result relies on the polinomiality of the tree-level anomalous dimensions as functions of $\Delta$. Otherwise one could expect higher transcendental weights at subleading orders.}. In particular, for the NLO term we will make the ansatz
\begin{align}
\nonumber
    h_i^{(1)}(\chi)&=s_{i0}(\chi)+s_{i1}(\chi)\log(\chi)+s_{i2}(\chi)\log(1-\chi)\,, \\
\label{NLO fi ansatz}    
    f_i^{(1)}(\chi)&=\tilde{s}_{i0}(\chi)+\tilde{s}_{i1}(\chi)\log(\chi)+\tilde{s}_{i2}(\chi)\log(1-\chi) \,,
\end{align}
where $s_{ij}$ and $\tilde{s}_{ij}$ are rational functions. Given that parts of the analytic bootstrap system are overconstrained, an oversight in the ansatz would lead to a vanishing solution. 

\subsubsection*{Crossing and braiding symmetry}
\label{Subsec: crossing and braiding}

The ansatz proposed in \eqref{NLO fi ansatz} is further constrained by crossing symmetry, which relates correlators with points 1 and 3 exchanged. In the OPE language, this corresponds to the two different OPE channels, the first is obtained through the OPE applied to operators at points 1 and 2 and the second is obtained from the OPE between operators at points 2 and 3. Crossing symmetry implies
\begin{align}
\nonumber
    {\bf h}(\chi,\rho_1,\rho_2) &= \left(\frac{\chi}{1-\chi}\right)\left(\frac{\rho_1}{1-\rho1}\right)^{-1} \left(\frac{\rho_2}{1-\rho_2}\right)^{-1} {\bf h}(1-\chi,1-\rho_1,1-\rho_2) \,, \\
\label{crossing f}
    {\bf f}(\chi,\rho_1) &= \left(\frac{\chi}{1-\chi}\right)^{2}\left(\frac{\rho_1}{1-\rho1}\right)^{-2}  {\bf f}(1-\chi,1-\rho_1) \,.
\end{align}
Equivalently, using the expansions \eqref{4-pt displ Ward} and \eqref{4-pt tilt Ward}, we get
\begin{align}
\nonumber
T_1&=0 \,, \\
\nonumber\chi \, h_1(1-\chi) - (1-\chi) \, h_1(\chi) &=0 \,, \\
\nonumber
(1-\chi) \, h_2 (\chi)+ \chi (h_1(1-\chi)+h_2(1-\chi))+T_2&=0 \,, \\
\nonumber
(1-\chi) \, f_1 (\chi) + \chi f_1 (1-\chi) + 2 D_0 &=0 \,, \\
\label{crossing-5}
(1-\chi)^2 f_2 (\chi)-\chi f_1 (1-\chi)- \chi^2 f_2 (1-\chi) +D_0 &=0 \,.
\end{align}

Additionally, there is another symmetry, known as \textit{braiding} symmetry \cite{Liendo:2018ukf}, that is inherited from the identity
\begin{align}
\label{braiding blocks}
    \left(\frac{\chi}{\chi-1}\right)^\Delta {}_2F(\Delta,\Delta,2\Delta,\frac{\chi}{\chi-1})=\left(-\chi\right)^\Delta {}_2F(\Delta,\Delta,2\Delta,\chi) \,,
\end{align}
satisfied by the conformal blocks. Let us focus for the moment on the leading-order term of ${\bf f}$, which was presented in Section \ref{sec LO}. It is interesting to note from \eqref{LO CFT data-7} that every block $G^\Delta_{r,\ell}$ (see eq. \eqref{conf block def}) that contributes to the leading-order OPE expansion of ${\bf f}$ has $\ell=0$ and satisfies that the quantum numbers $\{ \Delta, \frac{r}{2}\}$ are either both even or both odd. Therefore, using \eqref{braiding blocks} we see that each of those blocks satisfies
\begin{equation}
G^\Delta_{r,\ell} \left( \frac{\chi}{\chi-1}, \frac{\rho_1}{\rho_1-1}, \frac{\rho_2}{\rho_2-2} \right) =G^\Delta_{r,\ell} \left( \chi, \rho_1, \rho_2 \right) \,,
\end{equation}
at leading order, and consequently
\begin{equation}
{\bf f}^{(0)} \left( \frac{\chi}{\chi-1}, \frac{\rho_1}{\rho_1-1} \right) ={\bf f}^{(0)} \left( \chi, \rho_1 \right) \,.
\label{brading f0}
\end{equation}
When moving to next-to-leading order we need to take into account the perturbative corrections to the scaling dimensions of the unprotected multiplets. Assuming an all-order vanishing contribution of the OPE coefficients that are zero at leading order we get, using \eqref{braiding blocks}, that\footnote{A similar argument, used in \cite{Liendo:2018ukf,Ferrero:2021bsb} was justified by looking at $\mathbb{P}$-odd operators in \cite{Cavaglia:2023mmu} using the quantum spectral curve and finding them to be absent at strong coupling.}
\begin{equation}
{\bf f}^{(1)} \left( \frac{\chi}{\chi-1}, \frac{\rho_1}{\rho_1-1} \right) \bigg|_{\log(-\chi) \to \log(\chi)}={\bf f}^{(1)} \left( \chi, \rho_1 \right) \,.
\end{equation}
Similarly, for ${\bf h}$ we have
\begin{align}
\label{brading h0}
{\bf h}^{(0)} \left( \frac{\chi}{\chi-1}, \frac{\rho_1}{\rho_1-1} \right) &={\bf h}^{(0)} \left( \chi, \rho_1 \right) \,, \\
\label{bradingh1}
{\bf h}^{(1)} \left( \frac{\chi}{\chi-1}, \frac{\rho_1}{\rho_1-1} \right) \bigg|_{\log(-\chi) \to \log(\chi)} &={\bf h}^{(1)} \left( \chi, \rho_1 \right) \,.
\end{align}
It is easy to check that the braiding symmetry \eqref{brading f0} and \eqref{brading h0}, as well as crossing symmetry conditions
\eqref{crossing f}, are satisfied by the leading order expressions \eqref{LO f}.

In what follows, we will study in detail how crossing and braiding constrain the functions in our NLO ansatz  \eqref{NLO fi ansatz}.
Let us begin with the ${\bf h}$ function,  which describes the tilt four-point correlator. By inserting \eqref{4-pt tilt Ward} in \eqref{crossing f}, \eqref{brading h0} and \eqref{bradingh1} we obtain
\begin{align}
    T_i^{(0)}=T_i^{(1)}=0 \,,
\end{align}
for $i=1,2$, which is consistent with the LO result found in Section \ref{sec LO}. 

With a little hindsight we find convenient to reparametrize the rational functions of the NLO ansatz  \eqref{NLO fi ansatz} as
\begin{align}
    s_{10}(\chi) & =\frac{(2-\chi ) \, p_1(\chi)+\frac{\chi}{2} \, p_2(\chi)}{1-\chi } \,, 
    \quad &&
    s_{11}(\chi)=\frac{\chi}{1-\chi } (r_1(\chi) -\tfrac{1}{2}r_2(\chi)) \,,
   \nonumber\\ 
    s_{12}(\chi) & = r_1  (1-\chi ) - \frac{1}{2} r_2(1-\chi)\,, && \nonumber\\
    s_{20}(\chi) & =
    \frac{\chi}{\chi-1}
    p_2(\chi)\,, \quad && \ s_{21}(\chi) = \frac{\chi }{1-\chi } r_2(\chi) \,,
\nonumber\\
 s_{22}(\chi) & =-r_1(1-\chi )-\tfrac{1}{2} r_2(1-\chi )\,. &&
\end{align}
The advantage of this parametrization is the simplification of most of the braiding constraints, which become 
\begin{align}
\label{brading cond pi ri-1}
    p_1(\chi) - p_1 \left(\frac {\chi}{1-\chi}\right) = 0  \qquad \qquad p_2(\chi) - p_2 \left( \frac {\chi}{1-\chi}\right) & = 0  \\
    \label{brading cond pi ri-2}
   r_1(\chi) + r_1\left( \frac {\chi}{1-\chi}\right)=0  \qquad  \qquad r_2(\chi )- r_2\left(\frac{\chi }{\chi -1}\right) & =0 
   \\
   2 (\chi -1) \, r_1(1-\chi )-2 \chi  \, r_1(\chi )+ 2
\,r_2\left(\frac{1}{1-\chi }\right)-(\chi -1) \, r_2(1-\chi )+\chi  \, r_2(\chi )&=0 
\label{braideingleft1} \\
\!\!\!\!r_1\left(\frac{1}{1-\chi }\right)+(\chi -1) \, r_1(1-\chi )+ \frac12 r_2\left(\frac{1}{1-\chi }\right)+\frac12 (\chi -1) \, r_2(1-\chi )+\chi  \, r_2(\chi ) &=0 
\label{braideingleft2}
\end{align}
while the crossing conditions  impose that
\begin{align}
\label{crossingleft1}
-2 (\chi +1) \, p_1(1-\chi )-2 (\chi -2) \, p_1(\chi )+(\chi -1) \, p_2(1-\chi )+\chi  \, p_2(\chi ) &=0 \,, \\
\label{crossingleft2}
 2 (\chi +1) \, p_1(1-\chi )+(\chi -1) \, p_2(1-\chi )-2 \chi  \, p_2(\chi ) &=0 \,.
\end{align}

The four conditions in \eqref{brading cond pi ri-1} and \eqref{brading cond pi ri-2} imply that $p_1(\chi)$, $p_2(\chi)$ and $r_2(\chi)$ are braiding-symmetric and that $r_1(\chi)$ is braiding-antisymmetric. Therefore, we propose the expansion
\begin{align}
\nonumber 
    p_i(\chi)&= \sum_{k=M_{p_i}}^{N_{p_i}} p_{i,k} \, \left(\frac{\chi^2}{\chi-1}\right)^k \quad \text{for} \quad i=1,2, \\
  \nonumber %
  r_2(\chi) &= \sum_{k=M_{r_2}}^{N_{r_2}} r_{2,k}\left(\frac{\chi^2}{\chi-1}\right)^k , \\
    \label{brading symm ansatz h1-3}
    r_1(\chi) &= \left(\chi-\frac{\chi}{\chi-1}\right)\sum_{k=M_{r_1}}^{N_{r_1}} r_{1,k}\left(\frac{\chi^2}{\chi-1}\right)^k,
\end{align}
for some finite integers $M_{p_i},N_{p_i},M_{r_i}$ and $N_{r_i}$\footnote{The construction of the ansatz as an expansion in powers of $\frac{\chi^2}{\chi-1}$ will be crucial to obtain finite sums.}. The next step in the analytic bootstrap will involve applying the remaining braiding and crossing constraints and further consistency conditions to establish the upper and lower bounds for these sums, and to relate the coefficients of \eqref{brading symm ansatz h1-3} among them.

We now turn to the ${\bf f}$ function, which specifies the displacement four-point correlators. We find it convenient to use the following parametrization
\begin{align} 
    \tilde{s}_{10}(\chi) & =-\chi (1-2\chi)\, \tilde{p}_1(\chi)-2D_0, \quad && \tilde{s}_{11}(\chi )=\frac{\chi^3}{(1-\chi)^2}\, \tilde{r}_1(\chi),
    \nonumber
    \\ 
   \nonumber \tilde{s}_{12}(\chi ) & = -\frac{(1-\chi)^2}{\chi} \, \tilde{r}_1(1-\chi),
\\
    \tilde{s}_{20}(\chi ) & = \chi (1-2\chi) \, \tilde{p}_1(\chi)+\chi^2 \, \tilde{p}_2(\chi)+D_0, \quad && 
    \tilde{s}_{21}(\chi )=  - \frac{(2-\chi) \chi^3}{2(1-\chi)^3} \tilde{r}_1 (\chi)+ \frac{\tilde{r}_2(\chi)}{\chi^2}, 
    \nonumber
    \\
    \tilde{s}_{22}(\chi ) & = \frac{(1-\chi)^2}{2\chi} \, \tilde{r}_1(1-\chi) + \frac{\chi^2}{(1-\chi)^4} \, \tilde{r}_2(1-\chi) .
\end{align}
Then, crossing and braiding imply
\begin{align}
\label{brading cond tpi tri-1}
    \tilde{p}_1(\chi)&= \tilde{p}_1 \left( 1-\chi\right) \,, \; \; \quad \qquad \tilde{p}_2(\chi) = \tilde{p}_2 \left( 1-\chi\right) \,, \\
    \label{brading cond tpi tri-2}
   \tilde{r}_1(\chi)&=-\tilde{r}_1\left( \frac {\chi}{1-\chi}\right) \,, \qquad \tilde{r}_2(\chi )  =\tilde{r}_2\left(\frac{\chi }{\chi -1}\right) \,,
   \\
\label{br displ extra 1}
2 \, D_0 \,(\chi -1)^2&=   (2 \chi -1) (\chi -1)^3 \, \tilde{p}_1(\chi )+(\chi +1) \, \tilde{p}_1\left(\frac{\chi }{\chi -1}\right) \,,  \\
   \chi ^4 \, \tilde{r}_1(\chi ) &=(\chi -1)^4 \tilde{r}_1(1-\chi )-\tilde{r}_1\left(\frac{1}{1-\chi }\right) \,,  \\
(1-2 \chi) \, \tilde{p}_1(\chi )&=-\chi  \, \tilde{p}_2(\chi )+\frac{\chi}{(\chi-1)^4}  \, \tilde{p}_2\left(\frac{\chi }{\chi
   -1}\right)-\frac{D_0 (\chi -2)}{(\chi-1)^2} \nonumber \\
   &\quad -\frac{(\chi +1)}{(\chi-1)^4} \, \tilde{p}_1\left(\frac{\chi }{\chi -1}\right)  \,, \\
   \label{br displ extra 4}
2 \frac{(\chi -1)^4}{\chi} \, \tilde{r}_2(\chi )&=-(\chi -1)^6
   \, \tilde{r}_1(1-\chi )-(\chi -1) \, \tilde{r}_1\left(\frac{1}{1-\chi }\right) +\chi^4  \left(\chi ^2-3 \chi +2\right) \, \tilde{r}_1(\chi )\nonumber \\
   &\quad -2\chi ^3  \left[(\chi -1)^4 \, \tilde{r}_2\left(\frac{1}{1-\chi }\right)+\, \tilde{r}_2(1-\chi )\right] \,. 
\end{align}
The conditions \eqref{brading cond tpi tri-1} and \eqref{brading cond tpi tri-2} imply that $\tilde p_i$ are crossing-symmetric, $\tilde r_1$ is braiding-antisymmetric and $\tilde r_2$ is braiding-symmetric, which naturally guide us to propose
\begin{align}
\nonumber
    \tilde{p}_i(\chi)&= \sum_{k=M_{\tilde{p}_i}}^{N_{\tilde{p}_i}} \tilde{p}_{i,k} \, \left[\chi (1-\chi) \right]^k \quad \text{for} \quad i=1,2, \\
   \nonumber  \tilde{r}_2(\chi) &= \sum_{k=M_{\tilde{r}_2}}^{N_{\tilde{r}_2}} \tilde r_{2,k}\left(\frac{\chi^2}{\chi-1}\right)^k , \\
    \label{brading symm ansatz f1-3}
    \tilde{r}_1(\chi) &= \left(\chi-\frac{\chi}{\chi-1}\right)\sum_{k=M_{\tilde{r}_1}}^{N_{\tilde{r}_1}} \tilde{r}_{1,k}\left(\frac{\chi^2}{\chi-1}\right)^k.
\end{align}
We will later impose the remaining conditions \eqref{br displ extra 1}-\eqref{br displ extra 4} in order to constrain the above sums. But first we will find useful to discuss further consistency conditions to be satisfied by the ansatz. These will imply the consistency of the correlators with their OPE expansions and a mild growth of the anomalous dimensions at large $\Delta$.

\subsubsection*{OPE constraints and behaviour of anomalous dimensions}

The consistency of the CFT requires the ability to write a convergent OPE expansion. This will be constraining in several ways. First of all, the OPE controls the boundary behaviour of the functions. In particular, the fact that the conformal blocks introduced in \eqref{conf block def} are finite for $\chi \to 0$ implies
\begin{align}
\nonumber
    h_i (\chi) & \sim \mathcal{O}(1) \quad \text{for} \quad \chi \to 0,\\
    f_i (\chi) & \sim \mathcal{O}(1) \quad \text{for} \quad \chi \to 0,
\end{align}
for $i=1,2$. These constraints, combined with the crossing conditions presented in \eqref{crossing-5} give
\begin{align}
    h_i (\chi) & \sim \mathcal{O}[(\chi-1)^{-1}] \quad \text{for} \quad i=1,2, \nonumber\\
    f_1 (\chi) & \sim \mathcal{O}[(\chi-1)^{-1}],\nonumber \\
    f_2 (\chi) & \sim \mathcal{O}[(\chi-1)^{-2}],
\end{align}
for $\chi \to 1$.

Now we will introduce a dynamical constraint on the anomalous dimensions of exchanged long operators, and we will do it by requiring them to have a mild behaviour at large $\Delta$. Let us start by considering the N$^L$LO contribution $\gamma_{\Delta,r,\ell}^{(L)}$ to the anomalous dimensions of the long multiplets that are exchanged in \eqref{B1 OPE}. It has been suggested in the literature that the behaviour of the anomalous dimensions $\gamma_{\Delta,r,\ell}^{(L)}$ in the limit $\Delta \to \infty$ is related to the Lagrangian of the AdS$_2$ dual theory \cite{Heemskerk:2009pn,Fitzpatrick:2010zm}. In particular, the more irrelevant the interactions that contribute to the N$^L$LO order term of the four-point function, the bigger the growth of $\gamma_{\Delta,r,\ell}^{(L)}$ for large $\Delta$. To be more specific, it has been found that for the 1/2 BPS line defects of ${\cal N}=4$ SYM and ABJM \cite{Liendo:2018ukf,Bianchi:2020hsz,Ferrero:2023gnu}
\begin{align}
\gamma_{\Delta,r,\ell}^{(L)} \sim \Delta^{L+1}
   \qquad\text{for\ } {\Delta\rightarrow\infty} \,,
\end{align}
As we will discuss in Section \ref{Sec: Holographic description}, the interaction terms that contribute to the NLO order of the four point functions \eqref{4-pt displ primary} and \eqref{4 pt tilt primary} are at most quartic terms with four derivatives. Taking into account that this is behaviour of the interaction terms is the same as for the 1/2 BPS line defects of ${\cal N}=4$ SYM and ABJM \cite{Heemskerk:2009pn,Fitzpatrick:2010zm}, we will therefore impose
\begin{align}
\gamma_{\Delta,r,\ell}^{(1)} \sim \Delta^{2}
   \qquad\text{for\ } {\Delta\rightarrow\infty} \,,
\end{align}
for our NLO ansatz.

The advantage of imposing consistency with the OPE expansion and a mild behaviour for the anomalous dimensions is that they impose strong constraints on the upper and lower bounds of the \eqref{brading symm ansatz h1-3} and \eqref{brading symm ansatz f1-3} expansions. These, combined with the remaining crossing and braiding conditions discussed in \eqref{braideingleft1}-\eqref{crossingleft2}, \eqref{br displ extra 1} and \eqref{br displ extra 4}, leave only four unconstrained coefficients in the ansatz, as we will present in the next section.

\subsubsection*{Constrained NLO result}

When matching the constrained ansatz for ${\bf h}$ and ${\bf f}$ with OPE expansions introduced in \eqref{f ope 2} we obtain 
\begin{align}
   h_1^{(1)}(\chi) = & 3 \, c_{2,0,0}^{(1)} \left(\frac{1}{\chi-1}-\frac{\chi \log (\chi )}{(\chi -1)^2}-\frac{\log (1-\chi )}{\chi }\right) \,, \nonumber\\
   h_2^{(1)}(\chi) = &    {3 \, c_{2,0,0}^{(1)}} \left(-\frac{\chi}{\chi-1} +\frac{\chi\left(\chi ^2-2 \chi +2\right)}{(\chi-1)^2} \log (\chi ) - \chi \log (1-\chi )\right) \,,
   \nonumber\\
f_1^{(1)}(\chi) = & 
   -\frac{2}{3}\frac{(\chi -2)}{(\chi-1)}
  \left( \tilde{b}^{(1)}_{1,0}+ \tilde{b}^{(1)}_{2,0}\right)
-\frac{1}{3}\frac{\chi ^2(\chi -2)}{(\chi-1)^2} \left(\tilde{b}^{(1)}_{1,0}-2 \tilde{b}^{(1)}_{2,0}\right) 
\log (\chi)\nonumber
\\
& +\frac{1}{3} \frac{\left(\chi ^2-1\right) }{\chi}\left(\tilde{b}^{(1)}_{1,0}-2 \tilde{b}^{(1)}_{2,0}\right) \log (1-\chi )\,,
   \nonumber\\
f_2^{(1)}(\chi) = & 
\frac{\left(\chi ^2-\chi +1\right)}{5(\chi -1)^2} 
\left(12 \tilde{c}^{(1)}_{2,0,0}+5 \tilde{b}^{(1)}_{1,0}-4 \tilde{b}^{(1)}_{2,0}\right)
- \frac{1}{3(\chi -1)} \left(\tilde{b}^{(1)}_{1,0}+\tilde{b}^{(1)}_{2,0}\right) \\
  - &
 \frac{ \chi^2\log (\chi )}{30 (\chi-1)^3}\left(
3\left(12 \tilde{c}^{(1)}_{2,0,0}+5 \tilde{b}^{(1)}_{1,0}-4 \tilde{b}^{(1)}_{2,0}\right)\left(2 \chi ^2-5\chi +5\right)+5\left(\tilde{b}^{(1)}_{1,0}-2 \tilde{b}^{(1)}_{2,0}\right) (\chi-1)
\right)
 \nonumber\\
   + &
  \frac{ \log (1-\chi )}{30 \chi}\left(3\left(12 \tilde{c}^{(1)}_{2,0,0}+5 \tilde{b}^{(1)}_{1,0}-4 \tilde{b}^{(1)}_{2,0}\right)\left(2 \chi ^2+\chi +2\right)+5\left(\tilde{b}^{(1)}_{1,0}-2 \tilde{b}^{(1)}_{2,0}\right) (\chi+2)\right) \,.\nonumber
   \end{align}
Moreover, for the topological constants we get
\begin{align}
   T_1^{(1)}&=T_2^{(1)}=0 \,,\qquad
\label{top const 2}
D_0=\tilde{b}^{(1)}_{1,0}+\tilde{b}^{(1)}_{2,0}\,.
\end{align}
Let us note that the parameters unfixed by the bootstrap can be related to the normalization of the three-point functions discussed in \eqref{3ptdispl} as
\begin{equation}
\tilde{b}_{1,0}^{(1)} = - 4\sigma^2 \,.
\end{equation}
So far, we see that we have obtained a solution parametrized by four in principle independent OPE coefficients. However, at this point it is useful to consider the anomalous dimensions $\gamma^{(1)}_{\Delta,0,0}$ and $\tilde{\gamma}^{(1)}_{\Delta,0,0}$ of the exchanged ${\cal L}^{\Delta}_{0,0}$ supermultiplets.
\begin{align}
\nonumber
\gamma^{(1)}_{\Delta,0,0} &= -3 \Delta (\Delta+1) \, c^{(1)}_{2,0,0} \,, \\
\tilde{\gamma}^{(1)}_{\Delta,0,0} &=\frac{\Delta  (\Delta +1) \left(\frac{1}{5} \left(\Delta ^2+\Delta +4\right) \left(12  \tilde{c}^{(1)}_{2,0,0}+5 \tilde{b}^{(1)}_{1,0}-4 \tilde{b}^{(1)}_{2,0}\right)+2 \tilde{b}^{(1)}_{1,0}-4 \tilde{b}^{(1)}_{2,0}\right)}{4
   \left(\Delta ^2+\Delta -2\right)} \,.
\end{align}
Let us recall that the bulk CFT$_2$, which is dual to type IIB string theory in $AdS_3\times S^3\times T^4$, can be obtained as a limit of the CFT$_2$ which is dual to string theory in $AdS_3\times S^3\times S^3\times S^1$. Before taking the limit, 1/2 BPS line defects are invariant under the $D(2,1;\sin^2 \Omega)$ supergroup \cite{Correa:2021sky}, where $\Omega$ is a parameter that measures the relative size of the two $S^3$ spheres (this set up will be further discussed in the following section). One can recover the $PSU(1,1|2) \times SU(2)_A$  invariant line defects associated to the $AdS_3\times S^3\times T^4$ case by taking the limit $\Omega\to 0$ or $\Omega\to \frac{\pi}{2}$. Interestingly, in this limit the displacement supermultiplet of the $D(2,1;\sin^2 \Omega)$ invariant defect splits and gives rise to the displacement and tilt supermultiplets of the $PSU(1,1|2)\times SU(2)_A$ invariant defect. Therefore, it is reasonable to assume that the ${\cal L}^{\Delta}_{0,0}$ operators that are exchanged in both the tilt and displacement OPE (see \eqref{B1 OPE}) have the same anomalous dimensions, i.e.\footnote{The same conclusion could be reached by assuming that the anomalous dimensions of long operators are proportional to the eigenvalue of the quadratic Casimir, as proved for the case of the 1/2 BPS line of ${\cal N}=4$ SYM in \cite{Ferrero:2023gnu}.}
\begin{equation}
\label{anom dim assumption}
\gamma^{(1)}_{\Delta,0,0}= \tilde{\gamma}^{(1)}_{\Delta,0,0} \,.
\end{equation}
With this last assumption we get
\begin{align}
\nonumber
c_{2,0,0}^{(1)} &= \frac{1}{36} \left(\tilde{b}_{1,0}^{(1)}-2\tilde{b}_{2,0}^{(1)}\right) \,, \\
\tilde{c}_{2,0,0}^{(1)} &= \frac{1}{18} \left(-10 \tilde{b}_{1,0}^{(1)}+11 \tilde{b}_{2,0}^{(1)}\right) \,,
\end{align}
and therefore
\begin{align}
\nonumber 
 h_1^{(1)}(\chi)&= 
 \frac{ \left(\tilde{b}_{1,0}^{(1)}-2\tilde{b}_{2,0}^{(1)}\right)}{12}
 \left(\frac{1}{\chi-1}-\frac{\chi \log (\chi )}{(\chi -1)^2}-\frac{\log (1-\chi )}{\chi }\right) \,, \nonumber\\
   h_2^{(1)}(\chi) &= 
    \frac{ \left(\tilde{b}_{1,0}^{(1)}-2\tilde{b}_{2,0}^{(1)}\right)}{12}
   \left(-\frac{\chi}{\chi-1} +\frac{\chi\left(\chi ^2-2 \chi +2\right)}{(\chi-1)^2} \log (\chi ) - \chi \log (1-\chi )\right) 
   \,, \nonumber\\
   f_1^{(1)}(\chi) &= 
   -\frac{2}{3}\frac{(\chi -2)}{(\chi-1)}
  \left( \tilde{b}^{(1)}_{1,0}+ \tilde{b}^{(1)}_{2,0}\right)
-\frac{1}{3}\frac{\chi ^2(\chi -2)}{(\chi-1)^2} \left(\tilde{b}^{(1)}_{1,0}-2 \tilde{b}^{(1)}_{2,0}\right) 
\log (\chi)\nonumber
\\
& +\frac{1}{3} \frac{\left(\chi ^2-1\right) }{\chi}\left(\tilde{b}^{(1)}_{1,0}-2 \tilde{b}^{(1)}_{2,0}\right) \log (1-\chi )  \nonumber  \\
   \label{NLO final 4}
f_2^{(1)}(\chi)&= -\frac{\chi ^2 \left(\tilde{b}_{1,0}^{(1)}-2 \tilde{b}_{2,0}^{(1)}\right)+3 \tilde{b}_{2,0}^{(1)}\chi -3 \tilde{b}_{2,0}^{(1)}}{3 (\chi -1)^2}  
+\frac{\chi ^2 \left(\chi ^2-3 \chi +3\right) \left(\tilde{b}_{1,0}^{(1)}-2
   \tilde{b}_{2,0}^{(1)}\right) \log (\chi )}{3 (\chi -1)^3}\nonumber
\\
&\quad -\frac{\chi}{3}  \left(\tilde{b}_{1,0}^{(1)}-2 \tilde{b}_{2,0}^{(1)}\right) \log (1-\chi ) \,.
\end{align}

Let us remark that the final result from the analytic bootstrap depends only on two OPE coefficients, $\tilde{b}^{(1)}_{1,0}$ and $\tilde{b}^{(1)}_{2,0}$. We would like to emphasize that this statement strongly relies on the assumption \eqref{anom dim assumption}. We will later check this hypothesis with an holographic computation, providing an holographic interpretation for the coefficients that are not fixed by the bootstrap process. To facilitate the comparison we also present the bootstrap result for the four-point functions \eqref{rho4pt} and \eqref{varphi4pt} at NLO. Using the bootstrap solution in \eqref{d14} and \eqref{h3} we
get
\begin{align}
\langle \rho(t_1)\rho(t_2)&\rho(t_3)\rho(t_4)
\rangle^{(1)} =     \nonumber
\\
&   
\frac{2\tilde{b}^{(1)}_{2,0}}{\tau_{12}^4\tau_{34}^4} \frac{(1 - \chi + \chi^2)}{(1 - \chi)^4} (12-36\chi+25\chi^2+10\chi^3+25\chi^4-36\chi^5+12 \chi^6)\nonumber
\\&   
-\frac{4\tilde{b}^{(1)}_{1,0}}{\tau_{12}^4\tau_{34}^4} \frac{(1 - \chi + \chi^2)}{(1 - \chi)^4} (3-9\chi+7\chi^2+\chi^3+7\chi^4-9\chi^5+3 \chi^6)\nonumber
\\&   
-\frac{6\left(\tilde{b}^{(1)}_{1,0}-2\tilde{b}^{(1)}_{2,0}\right)}{\tau_{12}^4\tau_{34}^4}  
\frac{\chi^4}{(1-\chi)^5}
(2 - 6 \chi + 20 \chi^2 - 30 \chi^3 + 25 \chi^4 - 
 11 \chi^5 + 2 \chi^6)\log(\chi)\nonumber
 \\&   
-\frac{6\left(\tilde{b}^{(1)}_{1,0}-2\tilde{b}^{(1)}_{2,0}\right)}{\tau_{12}^4\tau_{34}^4}  
\frac{1}{\chi}
(2 - \chi  - 
 \chi^5 + 2 \chi^6)\log(1-\chi) \,,
\label{rho4ptbootstrap}
\end{align}
and
\begin{align}
\langle \varphi(t_1)\varphi(t_2)\varphi(t_3)\varphi(t_4)
\rangle^{(1)} = & -\frac{2\left(\tilde{b}^{(1)}_{1,0}-2\tilde{b}^{(1)}_{2,0}\right)}{3\tau_{12}^2\tau_{34}^2}\left(2\frac{(1-\chi+\chi^2)^2}{(1-\chi)^2}
+\frac{2-\chi-\chi^3+2\chi^4}{\chi}\log(1-\chi)\right.
\nonumber
\\
\ & + \left.
\frac{\chi^2(2 - 4 \chi + 9 \chi^2 - 7 \chi^3 + 2 \chi^4)}{(1-\chi)^3}
\log(\chi)\right) \,.
\label{varphi4ptbootstrap}
\end{align}

\subsubsection*{Extracting the CFT data}

By comparing the NLO solution \eqref{top const 2} and \eqref{NLO final 4} with the corresponding superconformal block expansions \eqref{f ope 2} we obtain that the NLO CFT data is
\begin{align}
\nonumber
\gamma^{(1)}_{\Delta,0,0}&=-\frac{\tilde{b}_{1,0}^{(1)}-2\tilde{b}_{2,0}^{(1)}}{12}\Delta (\Delta+1)\,, &\gamma^{(1)}_{\Delta,0,2}&=-\frac{\tilde{b}_{1,0}^{(1)}-2\tilde{b}_{2,0}^{(1)}}{12}\Delta (\Delta+1) \,, \\
\nonumber
 b_{1,0}^{(1)}&=0 \,, &b_{1,2}^{(1)}&=0 \,, \\
c_{\Delta,0,0}^{(1)}&=\partial_\Delta \left(\gamma^{(1)}_{\Delta,0,0}\,  c_{\Delta,0,0}^{(0)}\right) \,, & c_{\Delta,0,2}^{(1)}&=\partial_\Delta \left(\gamma^{(1)}_{\Delta,0,2} \, c_{\Delta,0,2}^{(0)}\right) \,,
\end{align}
and
\begin{align} \nonumber
\tilde{\gamma}^{(1)}_{\Delta,0,0}&=-\frac{1}{12} \Delta  (\Delta +1) \left(\tilde{b}_{1,0}^{(1)}-2 \tilde{b}_{2,0}^{(1)}\right)\,, \\
\nonumber
\tilde{\gamma}^{(1)}_{\Delta,2,0}&=-\frac{1}{12} (\Delta -1) (\Delta +2) \left(\tilde{b}_{1,0}^{(1)}-2 \tilde{b}_{2,0}^{(1)}\right) \,, \\ \nonumber
\tilde{c}_{\Delta,2,0}^{(1)}&= \partial_\Delta \left(\tilde{c}_{\Delta,2,0}^{(0)} \tilde{\gamma}_{\Delta,2,0}^{(1)}\right) +\frac{\tilde{b}_{1,0}^{(1)} \, \tilde{c}_{\Delta, 2,0}^{(0)}}{2 \Delta  (\Delta +1)}\,,
\\
\tilde{c}_{\Delta,0,0}^{(1)}&= \partial_\Delta \left(\tilde{c}_{\Delta,0,0}^{(0)} \tilde{\gamma}_{\Delta,0,0}^{(1)} \right)-\frac{3 \tilde{b}_{1,0}^{(1)} \, \tilde{c}_{\Delta,0,0}^{(0)}}{2 (\Delta -1) (\Delta +2)} \,.
\end{align}
Let us highlight two interesting limits in the above formulas by doing an informal presentation of what we might expect from the string theory perspective. Firstly, for $\tilde{b}_{1,0}^{(1)} \to 0$ the OPE coefficients take an expression analogous to the one presented in \cite{Bianchi:2020hsz, Giombi:2017cqn, Liendo:2018ukf},
suggesting that this limit corresponds with the pure R-R flux case, where the dual worldsheet is perpendicular to the boundary. On the other hand, when $\tilde{b}_{1,0}^{(1)} \to 2 \tilde{b}_{2,0}^{(1)}$ all the anomalous dimensions vanish. There is a particular limit where we might expect a somehow singular behaviour. This limit will correspond with  the pure NS-NS flux case, where the dual worldsheet becomes parallel to the boundary. In the next section we will present an holographic description that supports the interpretation of these two limits.

\section{Holographic description}
\label{Sec: Holographic description}

The BPS conformal lines we are studying in this article can also be described holographically, in terms of open strings in the type IIB supergravity background $AdS_3\times S^3\times T^4$, which can be obtain as a limit of $AdS_3\times S^3\times S^3\times S^1$. We will begin the holographic analysis with a review of the description of open strings and their quantum fluctuations in the latter background, and then we will take the $S^3\times S^1 \rightarrow T^4$ limit to compute Witten diagrams. The symmetry algebra of the $AdS_3\times S^3\times S^3\times S^1$ background is the direct product of two exceptional superalgebras $\mathfrak{d}(2, 1; \sin^2 \Omega) \otimes \mathfrak{d}(2, 1; \sin^2 \Omega)$ \cite{Gauntlett:1998kc,Babichenko:2009dk}.
Such background is described by the metric         
\begin{equation}
 ds^2 = L^2 ds^2(AdS_3)+\frac{L^2}{\sin^2\Omega} ds^2(S^3_+)+ \frac{L^2}{\cos^2\Omega} ds^2(S^3_-)+d\theta^2\,,
\end{equation}
where $\Omega$ is a parameter that sets the relative radii of the 3-spheres. Along with the metric, there is a constant dilaton $\Phi$ and a 3-form flux, which can be the Ramond-Ramond (R-R) one, the Neveu Schwarz-Neveu Schwarz (NS-NS) or a mixture of them,
\begin{align}
F_{(3)} & = d C_{(2)} =  -2e^{-\Phi} L^2 \cos\lambda
\left({\rm vol}(AdS_3)+\frac{1}{\sin^2\Omega} {\rm vol}(S^3_+)+ \frac{1}{\cos^2\Omega} {\rm vol}(S^3_-) \right)\,,
\\
H_{(3)} & = d B_{(2)} = 2 L^2 \sin\lambda
\left({\rm vol}(AdS_3)+\frac{1}{\sin^2\Omega} {\rm vol}(S^3_+)+ \frac{1}{\cos^2\Omega} {\rm vol}(S^3_-) \right)\,.
\end{align}
This introduces another parameter $\lambda$ which, taking values in the range $0\leq \lambda \leq \frac{\pi}{2}$, interpolates between pure R-R and  pure NS-NS cases. 

Supersymmetric open strings ending along a line on the boundary of these type of backgrounds were studied in \cite{Correa:2021sky}. In particular, certain open string forming an angle $\frac{\pi}{2}-\lambda$ with the boundary and sitting at fixed positions on the spheres were found to be 1/2 BPS.  Using Poincar\'e coordinates for the (Euclidean) $AdS_3$ space 
\begin{equation}
ds^2(AdS_3) =\frac{1}{z^2}\left(dt^2+dx^2+dz^2\right)\,,
    \label{poincare}
\end{equation}
this worldsheet configurations are parametrized by
\begin{equation}
t = \frac{\tau}{\cos \lambda}\,,\qquad
x = - \tan\lambda \ \sigma\,,\qquad 
z = \sigma\,.
\label{susyconfiguration}
\end{equation}
The induced metric on the worldsheet
\begin{equation}
ds^2 = \frac{L^2}{\sigma^2 \cos^2 \lambda }
\left(d\tau^2 + d\sigma^2\right)\,,
\end{equation}
describes an Euclidean $AdS_2$ of radius $R:=L/\cos \lambda$. The complete symmetry algebra of this classical configuration is $\mathfrak{d}(2, 1; \sin^2 \Omega)\subset \mathfrak{d}(2, 1; \sin^2 \Omega) \otimes \mathfrak{d}(2, 1; \sin^2 \Omega)$, whose bosonic subalgebra is 
$\mathfrak{sl}(2,\mathbb{R}) \oplus \mathfrak{su}(2) \oplus \mathfrak{su}(2)$. Thus, the set of fluctuations around this classical solution should be classified according the representations of this supergroup.

\subsection{Effective action}

To study the fluctuations around this classical configuration, we start from the Nambu-Goto
action for the bosonic coordinates of the string coupled to the NS-NS $B$-field
\begin{equation}
\label{bosonic action}
S_B= \frac{L^2}{2\pi\alpha'}\int d^2 \sigma \; \sqrt{h} + \frac{L^2}{4\pi\alpha'} \int d^2 \sigma \; \epsilon^{\alpha \beta} B_{\alpha\beta}\,.
\end{equation}
Here $h_{\alpha\beta}$ and $B_{\alpha\beta}$ are the induced metric and the pullback of the NS-NS $B$-field on the worldsheet, respectively. We have factored out $L^2$ in order to define $g:=\tfrac{L^2}{2\pi\alpha'}$, where $g^2$ is a parameter proportional to the 't Hooft coupling in the dual CFT$_2$.  Thus, we can implement a semiclassical expansion by taking the large $g$ limit and organizing terms according to their inverse powers of $g$. From now on
\begin{equation}
ds^2 = \frac{1}{z^2\cos^2\lambda }(d\tau^2+dz^2)\quad\Rightarrow\quad
h^{\alpha\beta} = \delta^{\alpha\beta} z^2 \cos^2\lambda \quad
\&\quad
 \sqrt{h} = \frac{1}{ z^2\cos^2\lambda} \,.
\end{equation}

The effective action for the fluctuations around the classical solution can be obtained making an expansion in terms of Riemann normal coordinates \cite{Drukker:2000ep,Forini:2015mca}. Let $X^{\mu}(q)$ be a geodesic such that $X^{\mu}(0)=X^{\mu}_{\sf cl}$. Using the geodesic equation we get
\begin{equation}
    \label{geodesic eq solution}
    \begin{aligned}
    X^{\mu}(q) &= X^{\mu}_{\sf cl}+ \frac{1}{g} q \zeta^{\mu}-\frac{1}{g^2}\frac{q^2}{2} \Gamma^{\mu}_{\nu \sigma} \zeta^{\nu} \zeta^{\sigma} + \frac{1}{g^3}\frac{q^3}{6} \left( 2 \Gamma^{\mu}_{\nu \kappa} \Gamma^{\kappa}_{\rho \sigma}-\partial_{\nu} \Gamma^{\mu}_{\rho \sigma} \right) \zeta^{\nu} \zeta^{\rho} \zeta^{\sigma} + \\
    &\quad + \frac{1}{g^4}\frac{q^4}{24} \bigg( 2 \Gamma^{\kappa}_{\rho \sigma} \partial_{\alpha} \Gamma^{\mu}_{\beta \kappa} + 2 \Gamma^{\mu}_{\alpha \kappa} \partial_{\beta} \Gamma^{\kappa}_{\rho \sigma}- 2 \Gamma^{\mu}_{\kappa \lambda} \Gamma^{\kappa}_{\alpha \beta} \Gamma^{\lambda}_{\rho \sigma} -4 \Gamma^{\mu}_{\alpha \kappa} \Gamma^{\kappa}_{\beta \lambda} \Gamma^{\lambda}_{\rho \sigma} - \partial_{\alpha} \partial_{\beta} \Gamma^{\mu}_{\rho \sigma} +  \\
  & \quad + \Gamma^{\kappa}_{\rho \sigma} \partial_{\kappa} \Gamma^{\mu}_{\alpha \beta}  + 2 \Gamma^{\kappa}_{\rho \sigma} \partial_{\alpha} \Gamma^{\mu}_{\kappa \beta} \bigg) \zeta^{\alpha} \zeta^{\beta} \zeta^{\rho} \zeta^{\sigma} + \mathcal{O}(\tfrac{1}{g^5})\,,
    \end{aligned}
\end{equation}
where the Christoffel symbols are to be evaluated in the classical solution $X^{\mu}_{\sf cl}$ and we have defined
\begin{equation}
    \label{rho coord riemann}
   \frac{1}{g} \zeta^{\mu}:=\left.\frac{dX^{\mu}}{dq}\right|_{q=0}\,.
\end{equation}
If we now consider $X^{\mu} = X^{\mu}(1)$, \eqref{geodesic eq solution} serves as an expansion of $X^\mu$ around $X^\mu_{\sf cl}$ in powers of a vector $\zeta^{\mu}$. Moreover, in order to get an expansion in terms of scalar fields, it is useful to define 
\begin{equation}
    \label{scalar phi}
    \phi^m= e^{m}_{\mu} \zeta^{\mu}\,,
\end{equation}
where $e^a_{\mu}$ are the inverse of the background vielbeins.

The effective action arises from substituting this expansion into \eqref{bosonic action}. Up to some boundary terms which can be discarded by appropriate counterterms, we get
\begin{equation}
\label{bosonic action expansion}
S_B= g S^{(0)}+S^{(2)}+\frac{1}{\sqrt g}S^{(3)}+ \frac{1}{g} S^{(4)}_{\phi_{\sf tr}}+ \frac{1}{g} S^{(4)}_{\phi^9}+\frac{1}{g} S^{(4)}_{S^3_+}+ \frac{1}{g} S^{(4)}_{S^3_-}+ \frac{1}{g} S^{(4)}_{\sf mix}
+{\cal O}(g^{-3/2})\,,
\end{equation}
where $S^{(0)}$ is the Nambu-Goto action evaluated in the classical solution and quadratic action for the 8 transverse modes is
\begin{align}
\label{bosonic action expansion-quadratic}
S^{(2)}&= \frac{1}{2} \int\! d^2 \sigma  \sqrt{h} \left[ \partial_{\alpha} \phi_{\sf tr} \partial^{\alpha} \phi_{\sf tr} + 2 \cos^2 \lambda \, \phi_{\sf tr}^2 +  \tfrac{1}{2} \partial_{\alpha} \phi^{a}_b \partial^{\alpha} \phi^b_a + \tfrac{1}{2} \partial_{\alpha} \phi^{\dot{a}}_{\dot{b}} \partial^{\alpha} \phi^{\dot{b}}_{\dot{a}} + \partial_{\alpha} \phi^9 \partial^{\alpha} \phi^9 \right] \,,
\end{align}
where we have introduced the notation $\phi_{\sf tr}$ for the transverse fluctuation in $AdS_3$, defined as
\begin{equation}
    \label{transverse fluctuation}
    \phi_{\sf tr} (\tau,\sigma):= \cos \lambda \; \phi^1(\tau,\sigma) +\sin \lambda\; \phi^2(\tau,\sigma)\,,
\end{equation}
and for the 3-spheres fluctuations we use 
\begin{align}
\label{definition phi su2+}
\phi^{a}_{b}  &:= \phi^{3} \left( \sigma^1 \right)^{a}_{b}  + \phi^{4} \left( \sigma^2 \right)^{a}_{b}  - \phi^{5} \left( \sigma^3 \right)^{a}_{b} \,, \\
\label{definition phi su2-}
\phi^{\dot{a}}_{\dot{b}} &:= \phi^{6} \left( \sigma^1 \right)^{\dot{a}}_{\dot{b}} + \phi^{7} \left( \sigma^2 \right)^{\dot{a}}_{\dot{b}} - \phi^{8} \left( \sigma^3 \right)^{\dot{a}}_{\rm \dot{b}}\,.
\end{align}

As expected, there is one massive fluctuation dual to an operator of $\Delta=2$ and seven massless fluctuations dual to operators of $\Delta=1$. Cubic and quartic interaction terms are the following\footnote{We would like to note that the one-loop correction to the partition function was computed in the $S^3 \times S^1 \to T^4$ limit in \cite{Pajer:2021bfr}.}
\begin{align}
S^{(3)} &= -\frac{2\sin\lambda}{3} \!\!\int\! \! d^2 \sigma  \sqrt{h} \left[  \cos^2\lambda\ \phi_{\sf tr}^3 + \tfrac{ \sin\Omega}{2} \epsilon^{\alpha\beta} \, \phi^a_b \partial_{\alpha} \phi^b_c \partial_{\beta}\phi^c_a + \tfrac{\cos\Omega}{2} \epsilon^{\alpha\beta} \,\phi^{\dot{a}}_{\dot{b}} \partial_{\alpha} \phi^{\dot{b}}_{\dot{c}} \partial_{\beta} \phi^{\dot{c}}_{\dot{a}} \right]
\label{qubicterms}
\\
S^{(4)}_{\phi^{\sf tr}} &=  \!\int\! \! d^2 \sigma  \sqrt{h} \left[  \tfrac{\cos^2\lambda}{3} \phi_{\sf tr}^4 - \tfrac{1}{8} \left( \partial_{\alpha} \phi_{\sf tr} \partial^{\alpha} \phi_{\sf tr} \right)^2 \right] \\
S^{(4)}_{\phi^{9}} &= -\frac{1}{8} \!\int\! \! d^2 \sigma  \sqrt{h}  \left( \partial_{\alpha} \phi^{9} \partial^{\alpha} \phi^{9} \right)^2 \label{S4phi9}\,, \\
S^{(4)}_{S^3_+} &= \int\! \! d^2 \sigma  \sqrt{h} \bigg[ \tfrac{1}{32} \left( \partial_{\alpha} \phi^{a}_b \partial^{\alpha} \phi^{b}_a \right)^2  - \tfrac{1}{16} \partial_{\alpha} \phi^{a}_b \partial_{\beta} \phi^{b}_a \partial^{\alpha} \phi^{c}_d \partial^{\beta} \phi^{d}_c \nonumber \\
&\qquad \qquad \quad \qquad \, \, -\tfrac{\sin^2 \Omega}{24} (  \phi^{a}_b  \phi^{b}_a \partial_{\alpha} \phi^{c}_d \partial^{\alpha} \phi^{d}_c - \phi^{a}_b \partial_{\alpha} \phi^{b}_a  \phi^{c}_d \partial^{\alpha} \phi^{d}_c ) \bigg]  \,, \\
S^{(4)}_{S^3_-} &= \!\int\! \! d^2 \sigma  \sqrt{h}\left[\tfrac{1}{32} \left( \partial_{\alpha} \phi^{\dot{a}}_{\dot{b}} \partial^{\alpha} \phi^{\dot{b}}_{\dot{a}} \right)^2  - \tfrac{1}{16} \partial_{\alpha} \phi^{\dot{a}}_{\dot{b}} \partial_{\beta} \phi^{\dot{b}}_{\dot{a}} \partial^{\alpha} \phi^{\dot{c}}_{\dot{d}} \partial^{\beta} \phi^{\dot{d}}_{\dot{c}} \right. \nonumber \\
&\qquad \qquad \quad \qquad \, \, - \left.\tfrac{\cos^2 \Omega}{24} (  \phi^{\dot{a}}_{\dot{b}}  \phi^{\dot{b}}_{\dot{a}} \partial_{\alpha} \phi^{\dot{c}}_{\dot{d}} \partial^{\alpha} \phi^{\dot{d}}_{\dot{c}} - \phi^{\dot{a}}_{\dot{b}} \partial_{\alpha} \phi^{\dot{b}}_{\dot{a}}  \phi^{\dot{c}}_{\dot{d}} \partial^{\alpha} \phi^{\dot{d}}_{\dot{c}} ) \right]  \,, \\
\label{bosonic action expansion-9}
S^{(4)}_{\sf mix} &= \!\int\! \! d^2 \sigma  \sqrt{h} \bigg[ \tfrac{1}{8} \left( \partial_{\alpha} \phi^{\sf tr} \partial^{\alpha} \phi^{\sf tr} \partial_{\beta} \phi^a_b \partial^{\beta} \phi^b_a + \partial_{\alpha} \phi^{\sf tr} \partial^{\alpha} \phi^{\sf tr} \partial_{\beta} \phi^{\dot{a}}_{\dot{b}} \partial^{\beta} \phi^{\dot{b}}_{\dot{a}} + 2 \, \partial_{\alpha} \phi^{\sf tr} \partial^{\alpha} \phi^{\sf tr} \partial_{\beta} \phi^9 \partial^{\beta} \phi^9 \right) 
\nonumber
\\
& \qquad \qquad  - \tfrac{1}{4} \left( \partial_{\alpha} \phi^{\sf tr} \partial_{\beta} \phi^{\sf tr} \partial^{\alpha} \phi^a_b \partial^{\beta} \phi^b_a + \partial_{\alpha} \phi^{\sf tr} \partial_{\beta} \phi^{\sf tr} \partial^{\alpha} \phi^{\dot{a}}_{\dot{b}} \partial^{\beta} \phi^{\dot{b}}_{\dot{a}} + 2 \, \partial_{\alpha} \phi^{\sf tr} \partial_{\beta} \phi^{\sf tr} \partial^{\alpha} \phi^9 \partial^{\beta} \phi^9 \right)   \nonumber \\
&\qquad \qquad  +\tfrac{1}{8} \left( \partial_{\alpha} \phi^{9} \partial^{\alpha} \phi^{9} \partial_{\beta} \phi^a_b \partial^{\beta} \phi^b_a + \partial_{\alpha} \phi^{9} \partial^{\alpha} \phi^{9} \partial_{\beta} \phi^{\dot{a}}_{\dot{b}} \partial^{\beta} \phi^{\dot{b}}_{\dot{a}} \right)  \nonumber \\
& \qquad \qquad - \tfrac{1}{4} \left( \partial_{\alpha} \phi^{9} \partial_{\beta} \phi^{9} \partial^{\alpha} \phi^a_b \partial^{\beta} \phi^b_a + \partial_{\alpha} \phi^{9} \partial_{\beta} \phi^{9} \partial^{\alpha} \phi^{\dot{a}}_{\dot{b}} \partial^{\beta} \phi^{\dot{b}}_{\dot{a}} \right)  \nonumber \\
&\qquad \qquad +\tfrac{1}{16} \partial_{\alpha} \phi^a_b \partial^{\alpha} \phi^b_a \partial_{\beta} \phi^{\dot{a}}_{\dot{b}} \partial^{\beta} \phi^{\dot{b}}_{\dot{a}} - \frac{1}{8} \partial_{\alpha} \phi^a_b \partial_{\beta} \phi^b_a \partial^{\alpha} \phi^{\dot{a}}_{\dot{b}} \partial^{\beta} \phi^{\dot{b}}_{\dot{a}} \bigg]
\,,
\end{align}

It is worth noting that the coupling with the NS-NS $B$-field gives rise to cubic interaction terms. This is novel compared to the case of open string fluctuations in $AdS_5\times S^5$, which does not exhibit cubic terms \cite{Drukker:2000ep,Giombi:2017cqn}. This is also different to the case of open string fluctuations in 
$AdS _4\times CP^3$, where the coupling with the NS-NS $B$-field only leads to boundary terms \cite{Bianchi:2020hsz,Correa:2023thy}.

In what follows, we will only consider the limit $\Omega\to\frac{\pi}{2}$ in which the background $AdS_3\times S^3\times S^3\times S^1$ becomes  $AdS_3\times S^3\times T^4$, and the symmetry algebra of the classical open string configuration $\mathfrak{d}(2,1;\sin^2\Omega)$ becomes
$\mathfrak{psu}(1,1,|2)\times \mathfrak{su}(2)_A$. In this limit, the transverse fluctuations accommodate in two different multiplets \cite{Correa:2021sky}. Fluctuations $\phi_{\sf tr}$ and  $\phi^a_{b}$ are part of the displacement multiplet, while the fluctuations   $\phi^{\dot a}_{\dot b}$ and $\phi^{9}$ are part of the tilt multiplet.

\subsection{Correlators from fluctuations $\phi_{\rm tr}$ }

From the massive scalar field $\phi_{\sf tr}$ we can compute correlation functions of the displacement operator $\rho$, the component of the displacement multiplet whose scale dimension is $\Delta = 2$ (see Section \ref{sec displ and tilt}). The corresponding bulk-to-boundary propagator is, in this case,
\begin{equation}
K_2(z,\tau,\tau') = {\cal C}_2 \left(\frac{z}{z^2+(\tau-\tau')^2}\right)^2\,,
\qquad
{\cal C}_2 = \frac{2}{3\pi}\,.
\end{equation}
The 2-point correlation function is simply
\begin{equation}
\langle {\cal O}_{\sf tr}(\tau_1){\cal O}_{\sf tr}(\tau_2)\rangle
= \frac{{\cal C}_2}{\tau_{12}^4}\,,\qquad\text{with\ } \tau_{ij}:= \tau_i-\tau_j
\end{equation}
and to have unit normalized 2-point functions one can define $\tilde{\cal O}_{\sf tr}={\cal O}_{\sf tr}/\sqrt{{\cal C}_2}$.

\begin{figure}[h!]
    \centering
    \begin{tikzpicture}[scale=0.8]
    \draw[thick] (0,0) circle [radius=0.8cm];
\draw[thick,cap=round,blue]  (-0.8,0) -- (0.8,0);
    \draw[thick,fill=white] (0.8,0) circle[radius=0.05cm];
    \draw[thick,fill=white] (-0.8,0) circle[radius=0.05cm];
     \node at (1.7,0) {+};
    \end{tikzpicture}
    \hspace{0.15cm}
    \begin{tikzpicture}[scale=0.8]
    \draw[thick] (0,0) circle [radius=0.8cm];
\draw[blue,thick,cap=round]  (-0.8,0) -- (0.8,0);    \draw[blue,fill=blue!25,thick] (0,0) circle [radius=0.2cm];
    \draw[thick,fill=white] (0.8,0) circle[radius=0.05cm];
    \draw[thick,fill=white] (-0.8,0) circle[radius=0.05cm];
    \end{tikzpicture}
    \hspace{0.5cm}
     \caption{Diagrammatic expansion of the 2-point function.}
    \label{fig:2pt}
\end{figure}
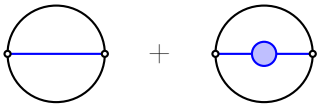

The 2-point function will also receive $\tfrac{1}{g}$ corrections from loop diagrams, as schematically 
depicted in Fig. \ref{fig:2pt}. These corrections are not explicitly computed, as they are absorbed into the normalization of the dual operator under the assumption that
\begin{equation}
\langle \tilde{\cal O}_{\sf tr}(\tau_1)\tilde{\cal O}_{\sf tr}(\tau_2)\rangle
= \frac{1}{\tau_{12}^4}\,,
\end{equation}
remains valid at all-loop order. In what follows we will omit the symbol $\ \tilde{}\ $ when referring to the unit normalized operators.

The three-point correlator is computed from the corresponding cubic vertex in \eqref{qubicterms}
\begin{align}
\langle {\cal O}_{\sf tr}(\tau_1){\cal O}_{\sf tr}(\tau_2){\cal O}_{\sf tr}(\tau_3)\rangle
= &\ 3!\frac{{\cal C}_2^{-3/2}}{\sqrt g} \frac{2}{3}\sin\lambda \int d^2 \sigma 
\frac{1}{z^2} K_2(z,\tau,\tau_1)K_2(z,\tau,\tau_2)K_2(z,\tau,\tau_3)
\nonumber
\\
= &\ \frac{4}{\sqrt g}\sin\lambda\ {\cal C}_2^{3/2} D_{2,2,1,1}(\tau_1,\tau_2,\tau_3,\tau_3)
= \frac{\frac{1}{\sqrt g}\sqrt{\frac{2}{3\pi}}\sin\lambda}{\tau_{12}^2\tau_{13}^2\tau_{23}^2}\,.
\end{align}
The conventions we use for the $D$-integrals are detailed in Appendix \ref{Dint}. From this result, we read the corresponding OPE coefficient 
\begin{equation}
C_{{\cal O}_{\sf tr}{\cal O}_{\sf tr}{\cal O}_{\sf tr}} = \frac{1}{\sqrt g}\sqrt{\frac{2}{3\pi}}\sin\lambda\,.
\label{Crho3}
\end{equation}

Up to a relative normalization \eqref{rhonorm}, ${\cal O}_{\sf tr}$ should be identified with the component $\rho$ of the displacement multiplet. This allow us to fix the value of the constant $\sigma$ in \eqref{3ptdispl}, which can be related to the OPE coefficient 
$\tilde{b}_{1,0}^{(1)}$  
\begin{equation}
    \sigma = \frac{1}{\sqrt g}\frac{1}{\sqrt{2\pi}}\sin\lambda
    \quad\Rightarrow\quad
    \tilde{b}_{1,0}^{(1)} = - 4\sigma^2 = -
    \frac{1}{g}\frac{2}{\pi}\sin^2\lambda \,.
\end{equation}
The leading order contribution of the four-point correlator is straightforwardly computed in terms of boundary-to-boundary propagators
\begin{align}
\langle {\cal O}_{\sf tr}(\tau_1){\cal O}_{\sf tr}(\tau_2) {\cal O}_{\sf tr}(\tau_3) {\cal O}_{\sf tr}(\tau_4)\rangle^{(0)}
= \frac{1}{\tau_{12}^4\tau_{34}^4}\left(1+\chi^4+\frac{\chi^4}{(1-\chi)^4}\right)\,.
\label{LOrho4}
\end{align}
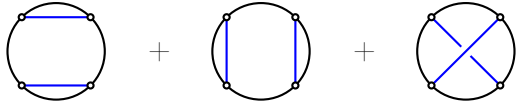
\begin{figure}[h!]
    \centering
    \begin{tikzpicture}[scale=0.8]
    \draw[thick] (0,0) circle [radius=0.8cm];
     \draw[blue,thick] (0.8*0.707,0.8*0.707)
     --(-0.8*0.707,0.8*0.707);
    \draw[blue,thick] (0.8*0.707,-0.8*0.707)
     --(-0.8*0.707,-0.8*0.707);
    \draw[thick,fill=white] (0.8*0.707,0.8*0.707) circle[radius=0.05cm];
    \draw[thick,fill=white] (0.8*0.707,-0.8*0.707) circle[radius=0.05cm];
    \draw[thick,fill=white] (-0.8*0.707,0.8*0.707) circle[radius=0.05cm];
    \draw[thick,fill=white] (-0.8*0.707,-0.8*0.707) circle[radius=0.05cm];
    \node at (1.7,0) {+};
    \end{tikzpicture}
    \hspace{0.15cm}
    \begin{tikzpicture}[scale=0.8]
    \draw[thick] (0,0) circle [radius=0.8cm];
     \draw[blue,thick] (0.8*0.707,0.8*0.707)
     --(0.8*0.707,-0.8*0.707);
     \draw[blue,thick] (-0.8*0.707, 0.8*0.707)
     --(-0.8*0.707,-0.8*0.707);
    \draw[thick,fill=white] (0.8*0.707,-0.8*0.707) circle[radius=0.05cm];
    \draw[thick,fill=white] (0.8*0.707,0.8*0.707) circle[radius=0.05cm];
    \draw[thick,fill=white] (-0.8*0.707,-0.8*0.707) circle[radius=0.05cm];
    \draw[thick,fill=white] (-0.8*0.707,0.8*0.707) circle[radius=0.05cm];
    \node at (1.7,0) {+};
    \end{tikzpicture}
\hspace{0.15cm}
 \begin{tikzpicture}[scale=0.8]
    \draw[thick] (0,0) circle [radius=0.8cm];
     \draw[blue,thick] (-0.8*0.707,0.8*0.707)
     --(0.8*0.707,-0.8*0.707);
     \draw[white,fill=white] (0,0) circle [radius=0.1cm];
     \draw[blue,thick] (-0.8*0.707, -0.8*0.707)
     --(0.8*0.707,0.8*0.707);
    \draw[thick,fill=white] (0.8*0.707,-0.8*0.707) circle[radius=0.05cm];
    \draw[thick,fill=white] (0.8*0.707,0.8*0.707) circle[radius=0.05cm];
    \draw[thick,fill=white] (-0.8*0.707,-0.8*0.707) circle[radius=0.05cm];
    \draw[thick,fill=white] (-0.8*0.707,0.8*0.707) circle[radius=0.05cm];
    \end{tikzpicture}
    \caption{Leading order contributions to the four-point function}
    \label{fig:4-ptLO}
 \end{figure}
 
At the next-to-leading order, we only need to consider connected diagrams. Diagrams like those shown in Fig. \ref{fig:4-ptLO}, where one of the propagators is corrected by a one-loop contribution, are absorbed into the normalization of the operator. The connected diagrams that contribute to the next-to-leading order include those with a single quartic vertex and exchange diagrams involving two cubic vertices in 3 different channels (see Fig. \ref{fig:4-ptNLO}).
\begin{align}
\langle {\cal O}_{\sf tr}(\tau_1){\cal O}_{\sf tr}(\tau_2) {\cal O}_{\sf tr}(\tau_3){\cal O}_{\sf tr}(\tau_4)\rangle^{(1)}
= {\sf q}_1 + {\sf q}_2 + {\sf ex} \,.
\label{NLOrho4}
\end{align}
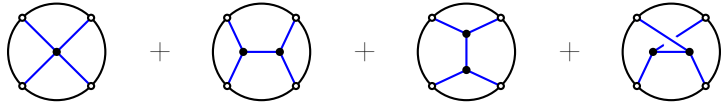
\begin{figure}[h!]
\centering
\begin{tikzpicture}[scale=0.8]
\draw[thick] (0,0) circle [radius=0.8cm];
\draw[blue,thick] (-0.8*0.707,0.8*0.707) --(0.8*0.707,-0.8*0.707);
\draw[blue,thick] (-0.8*0.707, -0.8*0.707)--(0.8*0.707,0.8*0.707);
\draw[thick,fill=white] (0.8*0.707,-0.8*0.707) circle[radius=0.05cm];
\draw[thick,fill=white] (0.8*0.707,0.8*0.707) circle[radius=0.05cm];
\draw[thick,fill=white] (-0.8*0.707,-0.8*0.707) circle[radius=0.05cm];
\draw[thick,fill=white] (-0.8*0.707,0.8*0.707) circle[radius=0.05cm];
\draw[thick,black,fill=black] (0,0) circle[radius=0.05cm];
\node at (1.7,0) {+};
\end{tikzpicture}
\hspace{0.15cm}
\begin{tikzpicture}[scale=0.8]
\draw[thick] (0,0) circle [radius=0.8cm];
\draw[blue,thick] (0.8*0.707,0.8*0.707) -- (0.3,0);
\draw[blue,thick] (0.8*0.707,-0.8*0.707) -- (0.3,0);
\draw[blue,thick] (-0.8*0.707,0.8*0.707) -- (-0.3,0);
\draw[blue,thick] (-0.8*0.707,-0.8*0.707) -- (-0.3,0);
\draw[blue,thick] (0.3,0) -- (-0.3,0);
\draw[thick,fill=white] (0.8*0.707,0.8*0.707) circle[radius=0.05cm];
\draw[thick,fill=white] (0.8*0.707,-0.8*0.707) circle[radius=0.05cm];
\draw[thick,fill=white] (-0.8*0.707,0.8*0.707) circle[radius=0.05cm];
\draw[thick,fill=white] (-0.8*0.707,-0.8*0.707) circle[radius=0.05cm];
\draw[thick,fill=black] (0.3,0) circle[radius=0.05cm];
\draw[thick,fill=black] (-0.3,0) circle[radius=0.05cm];
\node at (1.7,0) {+};
\end{tikzpicture}
\hspace{0.15cm}
\begin{tikzpicture}[scale=0.8]
\draw[thick] (0,0) circle [radius=0.8cm];
\draw[thick,blue] (0.8*0.707,0.8*0.707) -- (0,0.3);
\draw[thick,blue] (-0.8*0.707,0.8*0.707) -- (0,0.3);
\draw[thick,blue] (0.8*0.707,-0.8*0.707) -- (0,-0.3);
\draw[thick,blue] (-0.8*0.707,-0.8*0.707) -- (0,-0.3);
\draw[thick,blue] (0,0.3) -- (0,-0.3);
\draw[thick,fill=white] (0.8*0.707,0.8*0.707) circle[radius=0.05cm];
\draw[thick,fill=white] (0.8*0.707,-0.8*0.707) circle[radius=0.05cm];
\draw[thick,fill=white] (-0.8*0.707,0.8*0.707) circle[radius=0.05cm];
\draw[thick,fill=white] (-0.8*0.707,-0.8*0.707) circle[radius=0.05cm];
\draw[thick,fill=black] (0,0.3) circle[radius=0.05cm];
\draw[thick,fill=black] (0,-0.3) circle[radius=0.05cm];
\node at (1.7,0) {+};
\end{tikzpicture}
\hspace{0.15cm}
\begin{tikzpicture}[scale=0.8]
\draw[thick] (0,0) circle [radius=0.8cm];
\draw[blue,thick] (0.8*0.707,0.8*0.707) -- (-0.3,0);
\draw[fill=white,white] (0,0.15) circle[radius=0.12cm];    
\draw[thick,blue] (0.8*0.707,-0.8*0.707) -- (0.3,0);
\draw[thick,blue] (-0.8*0.707,0.8*0.707) -- (0.3,0);
\draw[thick,blue] (-0.8*0.707,-0.8*0.707) -- (-0.3,0);
\draw[thick,blue] (0.3,0) -- (-0.3,0);
\draw[thick,fill=white] (0.8*0.707,0.8*0.707) circle[radius=0.05cm];
\draw[thick,fill=white] (0.8*0.707,-0.8*0.707) circle[radius=0.05cm];
\draw[thick,fill=white] (-0.8*0.707,0.8*0.707) circle[radius=0.05cm];
\draw[thick,fill=white] (-0.8*0.707,-0.8*0.707) circle[radius=0.05cm];
\draw[thick,fill=black] (0.3,0) circle[radius=0.05cm];
\draw[thick,fill=black] (-0.3,0) circle[radius=0.05cm];
\end{tikzpicture}
\caption{Connected next-to-leading order contributions to the four-point function}
\label{fig:4-ptNLO}
\end{figure}

There are two types  of quartic vertices. From those without derivatives we have
\begin{align}
{\sf q}_1 \ &= - 4! \frac{1}{g}\frac{{\cal C}_2^{-2}}{3} \int d^2 \sigma 
\frac{1}{z^2} K_2(z,\tau,\tau_1)K_2(z,\tau,\tau_2)K_2(z,\tau,\tau_3)K_2(z,\tau,\tau_4)\\
\ & = -\frac{8}{g} {\cal C}_2^2 D_{2,2,2,2}(\tau_1,\tau_2,\tau_3,\tau_4) \,.
\end{align}
\noindent
The quartic vertex with derivatives gives
\begin{align}
{\sf q}_2 \ &= \frac{8 {\cal C}_2^{-2}}{g}\frac{\cos^2\lambda}{8} \left(\delta^{\alpha\beta}\delta^{\gamma\rho}+\delta^{\alpha\gamma}\delta^{\beta\rho}+\delta^{\alpha\rho}\delta^{\beta\gamma}\right)
 \nonumber \\
& \hspace{1cm} \int d^2 \sigma \, z^2 \, \partial_\alpha K_2(z,\tau,\tau_1)\partial_\beta K_2(z,\tau,\tau_2) \partial_\gamma K_2(z,\tau,\tau_3) \partial_\rho K_2(z,\tau,\tau_4)\,,
\end{align}
which, using the identity \eqref{dKdK} in the appendix \ref{Dint}, becomes
\begin{align}
{\sf q}_2  =\ & \frac{\cos^2\lambda}{g}16{\cal C}_2^2(D_{2,2,2,2}-2\tau_{12}^2 D_{3,3,2,2}-2\tau_{34}^2 D_{2,2,3,3}+4\tau_{12}^2\tau_{34}^2 D_{3,3,3,3}) 
 \nonumber \\
& +\frac{\cos^2\lambda}{g}16{\cal C}_2^2(D_{2,2,2,2}-2\tau_{13}^2 D_{3,2,3,2}-2\tau_{24}^2 D_{2,3,2,3}+4\tau_{13}^2\tau_{24}^2 D_{3,3,3,3}) 
\nonumber\\
& +\frac{\cos^2\lambda}{g}16{\cal C}_2^2(D_{2,2,2,2}-2\tau_{14}^2 D_{3,2,2,3}-2\tau_{23}^2 D_{2,3,3,2}+4\tau_{14}^2\tau_{23}^2 D_{3,3,3,3}) \,.
 \end{align}
The exchange diagrams are
\begin{align}
{\sf ex}  =\ & \frac{16\sin^2\lambda}{g} {\cal C}_2^2
\, \frac{\chi^4}{\tau_{12}^4\tau_{34}^4}\, \left(f_{\sf ex}^{(12)}+f_{\sf ex}^{(13)}+f_{\sf ex}^{(14)}\right)
\,.
 \end{align}
These exchange diagrams can be related to simpler contact integrals through a differential operator \cite{DHoker:1999mqo,DHoker:1999kzh} (see \cite{Zhou:2018sfz,Bliard:2022xsm} for recent reviews) :
\begin{equation}
(C_{(2)}^{ij} - 2)f^{(ij)}_{\sf ex}=  D_{2,2,2,2}\tau_{13}^4\tau_{24}^4
\end{equation}
where $C_{(2)}^{ij}$ is the quadratic Casimir acting on the two boundary points $i$ and $j$.
Some of the integration constants in $f^{(ij)}$ are fixed imposing crossing and braiding symmetry
\begin{align}
f^{(12)}_{\sf ex}(\chi) =&\ f^{(14)}_{\sf ex}(1-\chi)\,,
\\
f^{(13)}_{\sf ex}(\chi) =&\ f^{(13)}_{\sf ex}(1-\chi)\,,
\\
f^{(13)}_{\sf ex}(\chi) =&\ (1-\chi)^{-4}f^{(14)}_{\sf ex}\left(\tfrac{\chi}{\chi-1}\right)\,.
\end{align} 
Then, we obtain
\begin{align}
{\sf ex}  =\ & \frac{1}{g}\frac{2\sin^2\lambda}{9 \pi\tau_{12}^4\tau_{34}^4}\frac{(1 - \chi + \chi^2)}{(1 - \chi)^4}
 (36 - 108 \chi + 73 \chi^2 + 34 \chi^3 + 73\chi^4 - 
    108 \chi^5 + 36 \chi^6)\nonumber\\
& + \frac{k}{g}\frac{\sin^2\lambda}{6\pi \tau_{12}^4\tau_{34}^4}\frac{(1 - \chi + \chi^2)}{(1-\chi)^4}
 \left( 1 - 3 \chi + 2 \chi^2 + \chi^3 + 2 \chi^4 - 3 \chi^5 + \chi^6\right)\nonumber
 \\
 & + \frac{1}{g}\frac{2\sin^2\lambda}{9 \pi\tau_{12}^4\tau_{34}^4}\frac{\chi^4}{(1 - \chi)^5}
 (31 - 93 \chi + 343 \chi^2 -531 \chi^3 + 448\chi^4 - 
    198 \chi^5 + 36 \chi^6)\log(\chi)\nonumber
 \\
 & + \frac{1}{g}\frac{3 k \sin^2\lambda}{ \pi\tau_{12}^4\tau_{34}^4}\frac{\chi^4}{(1 - \chi)^5}
 (2 - 6 \chi + 20 \chi^2 -30 \chi^3 + 25\chi^4 - 
    11 \chi^5 + 2 \chi^6)\log(\chi)\nonumber
    \\
    & +  \frac{1}{g}\frac{\sin^2\lambda}{ 9\pi\tau_{12}^4\tau_{34}^4}\frac{1}{\chi}
    (72 - 36 \chi -4 \chi^2 -2 \chi^3 -4\chi^4 - 
    36 \chi^5 + 72 \chi^6)\log(1-\chi)\nonumber
    \\
 & + \frac{1}{g}\frac{3 k \sin^2\lambda}{ \pi\tau_{12}^4\tau_{34}^4}\frac{1}{\chi}(2-\chi-\chi^5+2\chi^6)
\log(1-\chi)\,,
    \end{align}
where $k$ is an integration constant, which can be fixed by considering our computation in the small $\chi$ limit. Collecting all the contributions to \eqref{LOrho4} and \eqref{NLOrho4},
we have
\begin{equation}
    \langle{\cal O}_{\sf tr}(\tau_1){\cal O}_{\sf tr}(\tau_2){\cal O}_{\sf tr}(\tau_3){\cal O}_{\sf tr}(\tau_4)
\rangle = 
\frac{1}{\tau_{12}^4\tau_{34}^4}
\left(1-\frac{1}{g} \frac{k\sin^2\lambda}{2\pi} \chi^2+ {\cal O}(\chi^3)\right) \,.
\end{equation}
Taking the same limit in the
conformal block expansion,
\begin{align}
\langle{\cal O}_{\sf tr}(\tau_1){\cal O}_{\sf tr}(\tau_2){\cal O}_{\sf tr}(\tau_3){\cal O}_{\sf tr}(\tau_4)
\rangle = &\  
\frac{1}{\tau_{12}^4\tau_{34}^4} \sum_{\Delta}
 C_{{\cal O}_{\sf tr}{\cal O}_{\sf tr}{\cal O}_{\Delta}}^2 
 g_\Delta(\chi)
\nonumber
\\
= & \  \frac{1}{\tau_{12}^4\tau_{34}^4}(1+\chi^2 C_{{\cal O}_{\sf tr}{\cal O}_{\sf tr}{\cal O}_{\sf tr}}^2+ {\cal O}(\chi^3)) \,.
\end{align}
Comparing the two expansion and using \eqref{Crho3} we determine that $k=-\frac{4}{3}$.

The final result for the next-to-leading order is
\begin{align}
\langle {\cal O}_{\sf tr}(\tau_1)&{\cal O}_{\sf tr}(\tau_2) {\cal O}_{\sf tr}(\tau_3){\cal O}_{\sf tr}(\tau_4)\rangle^{(1)}
= \nonumber \\
&   
-\frac{1}{g}\frac{1}{3 \pi\tau_{12}^4\tau_{34}^4} \frac{(1 - \chi + \chi^2)}{(1 - \chi)^4} (12-36\chi+25\chi^2+10\chi^3+25\chi^4-36\chi^5+12 \chi^6)\nonumber
\\&   
+\frac{1}{g}\frac{\sin^2\lambda}{ \pi\tau_{12}^4\tau_{34}^4} \frac{(1 - \chi + \chi^2)}{(1 - \chi)^4} (4-12\chi+9\chi^2+2\chi^3+9\chi^4-12\chi^5+4 \chi^6)\nonumber
\\&   
-\frac{2}{g}\frac{\cos^2\lambda}{\pi\tau_{12}^4\tau_{34}^4}
\frac{\chi^4}{(1-\chi)^5}
(2 - 6 \chi + 20 \chi^2 - 30 \chi^3 + 25 \chi^4 - 
 11 \chi^5 + 2 \chi^6)\log(\chi)\nonumber
 \\&   
-\frac{2}{g}\frac{\cos^2\lambda}{\pi\tau_{12}^4\tau_{34}^4}  
\frac{1}{\chi}
(2 - \chi  - 
 \chi^5 + 2 \chi^6)\log(1-\chi) \,.
\label{NLOrho4final}
\end{align}

Multiplying by 9, to take into account the different normalization we used for the displacement operator $\rho$, \eqref{NLOrho4final} matches exactly the expression \eqref{rho4ptbootstrap} obtained with bootstrap, provided that
\begin{equation}
\tilde{b}_{1,0}^{(1)}  =      
-\frac{1}{g}\frac{2}{\pi}\sin^2\lambda\,,
\qquad
\tilde{b}_{2,0}^{(1)} = -\frac{1}{g}\frac{1}{2\pi}\left(3-\sin^2\lambda\right)\,.
\label{coeffidentification}
\end{equation}

As we have anticipated, the $\tilde{b}_{1,0}^{(1)}\to 0$ limit corresponds to the case in which the flux of the background is pure R-R. The other critical limit we had identified in the bootstrap analysis
was $\tilde{b}_{1,0}^{(1)}\to 2\tilde{b}_{2,0}$, in which all the anomalous dimensions of long multiplets become vanishing. We can see now that this is not a physically interesting limit in our setup, given that $\tilde{b}_{1,0}^{(1)}-
2\tilde{b}_{2,0}^{(1)}\propto \cos^2\lambda$ becomes vanishing for $\lambda\to\frac{\pi}{2}$, i.e. when the flux of the background is pure NS-NS. In this limit the worldsheet becomes parallel to the boundary of AdS$_3$ and no longer describes a line defect in a CFT$_2$.

\subsection{Correlators from  fluctuations $\phi^{9}$ }

In this case, the massless scalar field is in correspondence with a descendant of the tilt multiplet, whose scale
dimension is $\Delta = 1$. The bulk-to-boundary propagator in this case is
\begin{equation}
K_1(z,\tau,\tau') = {\cal C}_1 \frac{z}{z^2+(\tau-\tau')^2}\,,
\qquad
{\cal C}_1 = \frac{1}{\pi}\,.
\end{equation}
From the vertex \eqref{S4phi9} and using the identity \eqref{dKdK}, the connected 4-point correlator (for unit normalized operators) becomes
\begin{align}
\langle {\cal O}_{9}(\tau_1){\cal O}_{9}(\tau_2) {\cal O}_{9}(\tau_3) {\cal O}_{9}(\tau_4)\rangle
 &=  \frac{\cos^2\lambda}{g\pi^2}(D_{1,1,1,1}-2\tau_{12}^2 D_{2,2,1,1}-2\tau_{34}^2 D_{1,1,2,2}+4\tau_{12}^2\tau_{34}^2 D_{2,2,2,2}) 
 \nonumber \\
& +\frac{\cos^2\lambda}{g\pi^2}(D_{1,1,1,1}-2\tau_{13}^2 D_{2,1,2,1}-2\tau_{24}^2 D_{1,2,1,2}+4\tau_{13}^2\tau_{24}^2 D_{2,2,2,2}) 
\nonumber\\
& +\frac{\cos^2\lambda}{g\pi^2}(D_{1,1,1,1}-2\tau_{14}^2 D_{2,1,1,2}-2\tau_{23}^2 D_{1,2,2,1}+4\tau_{14}^2\tau_{23}^2 D_{2,2,2,2}) 
\nonumber\\
\ &= \frac{\cos^2\lambda}{\pi g \tau_{12}^2\tau_{34}^2} 
\left(2\frac{(1-\chi+\chi^2)^2}{(1-\chi)^2}
+\frac{2-\chi-\chi^3+2\chi^4}{\chi}\log(1-\chi)\right.
\nonumber\\
\ & \hspace{1.85cm}+ \left.
\frac{\chi^2(2 - 4 \chi + 9 \chi^2 - 7 \chi^3 + 2 \chi^4)}{(1-\chi)^3}
\log(\chi)\right) 
\label{O94ptwitten}
\end{align}

To compare with the bootstrap result \eqref{varphi4ptbootstrap}, we must multiply \eqref{O94ptwitten} by 4, as the operator $\hat\varphi$ has a norm $\sqrt2$. With the identification of coefficients given in \eqref{coeffidentification}, the agreement with \eqref{varphi4ptbootstrap} is exact.

\section{Conclusions}
\label{sec conclusions}

In this paper we have applied analytic bootstrap techniques to study 1/2 BPS line defects in the CFT$_2$ which is dual to type IIB string theory in $AdS_3 \times S^3 \times T^4$ with mixed R-R and NS-NS three-form flux. In particular, we have studied two-, three- and four-point functions of the line defects excitations contained in the displacement and tilt supermultiplets, performing the analysis up to next-to-leading order in the strong-coupling expansion. We have found that supersymmetry completely fixes the two-point and three-point functions at most up to a single parameter. As for the four-point functions, the analytic bootstrap determines the full result up to two coefficients corresponding to the parameters in three-point functions. 
From these results we have obtained CFT data associated to the line defect, up to next-to-leading order and expressed in terms of these two parameters that are not fixed by the bootstrap procedure. We have checked that the bootstrap results perfectly match the holographic expectations coming from the Witten diagram expansion of the correlators. This holographic description allowed us to obtain an interpretation of the two coefficients that parametrize the bootstrap result, relating them to the string tension, the AdS radius and the tilt angle of the dual string. 

Our results lead to many interesting follow-up questions. On the one hand, it would be interesting to pursue the bootstrap analysis beyond next-to-leading order in the strong coupling expansion. At these orders we start to get loop Witten diagrams, which are fiendishly difficult \cite{Carmi:2018qzm} for direct integration methods. Therefore, the analytic bootstrap appears as a very appealing method to obtain further subleading corrections to the correlators. To compute these, the knowledge of the protected three-point functions then becomes indispensable. In the case of $\mathcal{N}=4$ SYM, the exact OPE coefficients were computed using localisation \cite{Giombi:2018qox}, and in the case of ABJM the knowledge of the bremsstrahlung function, also known from localisation \cite{Bianchi:2018scb,Bianchi:2014laa,Bianchi:2017ozk,Correa:2014aga}, was  sufficient. The corresponding exact computation of the OPE coefficients of protected operators would be necessary for a self-sufficient analytic bootstrap computation of the next subleading orders. The computation of the CFT data at higher-orders would also motivate a degeneracy analysis for the operators exchanged in the OPE. A detailed discussion of the mixing problem for the 1/2 BPS line defect of ${\cal N}=4$ sYM can be found in \cite{Ferrero:2023znz,Ferrero:2023gnu}.

On the other hand, it is a natural question to understand how far we might push the bootstrap program as we reduce the number of supersymmetries. In this regard, we should make the distinction about the number of supersymmetries preserved by the line defect and the number of supersymmetries preserved by the bootstrapped multiplet. Concerning the former, it would be interesting to enquire  whether analytic conformal bootstrap techniques can be applied to study the bosonic 1/6 BPS Wilson loop of ABJM \cite{Drukker:2008zx}, which preserves only 2 supercharges. As for the bootstrap of less supersymmetric supermultiplets, one could consider the case of the displacement supermultiplet associated to the 1/2 BPS line defect of the CFT$_2$ which is dual to type IIB string theory in $AdS_3 \times S^3 \times S^3 \times S^1$. Notably, this multiplet is just 1/4 BPS, an therefore is left invariant by only 1 supercharge. 
The correlators of these less supersymmetric multiplets should depend on an additional parameter that accounts for the relative size of the two $S^3$. Interestingly, in the limit $S^3 \times S^1 \rightarrow T^4$ this supermultiplet splits into two smaller multiplets \cite{Correa:2023lsm}, which are precisely the displacement and tilt supermultiplets that we have studied in this paper. The study of four-point correlator of the displacement multiplet in the $AdS_3 \times S^3 \times S^3 \times S^1$ case would naturally provide the mixed correlators between two displacement and two tilt operators in the limit $S^3 \times S^1 \rightarrow T^4$.

Furthermore, the bootstrap of the 1/4 BPS displacement multiplet in the $AdS_3 \times S^3 \times S^3 \times S^1$ case opens another interesting new question. Let us note that in our analysis a crucial step to constrain the superconformal blocks and four-point functions was the existence of a topological sector.  To construct it we used the fact that our tilt and displacement multiplets were annihilated by two supercharges. Then, the R-symmetry rotating those supercharges could be combined with time translations to build the topological translations. This does not seem to be possible if the bootstrapped multiplets were annihilated by only 1 supercharge, which is in fact what happens with the 1/4 BPS displacement multiplet in the $AdS_3 \times S^3 \times S^3 \times S^1$ case. We believe it might be interesting to explore the possibility of applying analytic bootstrap methods without the existence of a topological sector. In particular, a construction of the superconformal blocks using the superconformal quadratic and cubic Casimir in the spirit of \cite{Cornagliotto:2017dup} could solve the dependency on nilpotent superconformal cross-ratios.

It would also be interesting to explore the Lagrangian description discussed in \cite{OhlssonSax:2014jtq} for the bulk CFT$_2$ in the limit of pure R-R flux, and to see if a Wilson line representation could be constructed for the 1/2 BPS line defects we have studied. This could open the door to a supersymmetric localisation analysis of these line defects, which might enable the exact determination of the bremsstrahlung function. These results could then be compared with the first strong-coupling orders of the bremsstrahlung function, that could be obtained from the four-point correlators presented here using integral formulas as in \cite{Drukker:2022pxk}. 

Finally, another interesting direction is the bootstrap of multi-point correlators. These types of correlators have recently gathered interest \cite{Barrat:2021tpn,Bliard:2023zpe,Giombi:2023zte}, and many of the tools used in this paper are applicable when extended to more than four points. In higher-point cases the conformal blocks become generalized Appell functions \cite{Rosenhaus:2018zqn}, the ansatz contains more complicated polylogarithms, and the Ward identities \cite{Bliard:2024und} play an important role. In particular, the six- and eight-point correlators contain CFT data that gives information about the spectrum of more generic long operators, needed to unmix CFT data as in \cite{Ferrero:2023znz}.

\section*{Acknowledgements}

We would like to thank Lorenzo Bianchi, Aleix Gimenei-Grau and Philine Van Vliet for useful discussions. The authors are grateful to Max Planck Institute for Physics in Munich for its hospitality. This work was partially supported by  PICT 2020-03749, PIP 02229, UNLP X910. DHC would like to acknowledge support from the ICTP through the Associates Programme (2020-2025). ML is supported by fellowships from CONICET (Argentina) and DAAD (Germany).
GB was funded by the European Union (ERC, FUNBOOTS, project number 101043588). ISL would like to acknowledge support from the ICTP through the Associates Programme (2023-
2028).  Views and opinions expressed are however those of the author only and do not necessarily reflect those of the European Union or the European Research Council. Neither the European Union nor the granting authority can be held responsible for them.

\appendix

\section{AdS$_2$ propagators and  $D$-integrals}
\label{Dint}
In this appendix, we review the conventions used for the AdS$_2$ propagators and the resulting $D$-integrals, which appear in the computation of certain tree-level Witten diagrams \cite{Freedman:1998tz,DHoker:1999kzh,Fitzpatrick:2011ia,Penedones:2010ue}. For a scalar field bulk-to-boundary propagator, we 
consider
\begin{equation}
K_\Delta(z,t,t') := {\cal C}_\Delta \bar K_\Delta(z,t,t') = {\cal C}_\Delta \left(\frac{z}{z^2+(t-t')^2}\right)^\Delta\,,
\end{equation}
where
\begin{equation}
{\cal C}_\Delta = \frac{\Gamma(\Delta)}{2\sqrt{\pi}\Gamma(\Delta+\tfrac{1}{2})}\,.
\end{equation}
For the scalar field bulk-to-bulk propagator, we have
\begin{equation}
G_\Delta(z,z',t,t') =
{\cal C}_\Delta (2u)^{-1}{}  _2F_1(\Delta,\Delta,2\Delta,2 u^{-1})\,,
\qquad
u = \frac{(z-z')^2+(t-t')^2}{2 z z'}\,.
\end{equation}

Tree-level Witten diagrams with a single vertex are computed using the so-called $D$-integrals. For quartic vertices, one defines
\begin{equation}
D_{\Delta_1,\Delta_2,\Delta_3,\Delta_4}(t_1,t_2,t_3,t_4)=
\int\frac{d^2\sigma}{z^2}
\prod_{i=1}^4 \bar K_{\Delta_i}(z,t,t_i)\,.
\end{equation}
The following identity,
\begin{align}
z^2\delta^{\alpha\beta}&\partial_{\alpha}\bar K_{\Delta_1}(z,t,t_1)\partial_{\beta}\bar K_{\Delta_2}(z,t,t_2)
\label{dKdK}
\\
&=\Delta_1\Delta_1\left(
\bar K_{\Delta_1}(z,t,t_1)\bar K_{\Delta_2}(z,t,t_2)-2(t_2-t_1)^2
\bar K_{\Delta_1+1}(z,t,t_1)\bar K_{\Delta_2+1}(z,t,t_2)
\right)\,,\nonumber
\end{align}
would be useful when dealing with integrals from vertices involving derivatives.

We can specify the $D$-integrals in terms of the functions of the cross-ratio associated to them,
\begin{equation}
D_{\Delta_1,\Delta_2,\Delta_3,\Delta_4}
=\frac{\pi^{d/2}\Gamma\left(\Sigma-\tfrac{d}{2}\right)}{2\Gamma(\Delta_1)\Gamma(\Delta_2)\Gamma(\Delta_3)\Gamma(\Delta_4)}\frac{t_{14}^{2\Sigma-2\Delta_1-2\Delta_4}t_{34}^{2\Sigma-2\Delta_3-2\Delta_4}}{t_{13}^{2\Sigma-2\Delta_4}t_{24}^{2\Delta_2}}
\bar D_{\Delta_1,\Delta_2,\Delta_3,\Delta_4}(\chi) \,,
\end{equation}
where $\Sigma = \tfrac{1}{2}(\Delta_1+\Delta_2+\Delta_3+\Delta_4)$. 

The $\bar{D}$-integrals needed for our computations are
\begin{align}
\bar D_{1,1,1,1}(\chi) = &\ -2 \frac{\log\chi}{1-\chi}-2 \frac{\log(1-\chi)}{\chi}\,,
\\
\bar D_{1,1,2,2}(\chi) = &\ \frac{1}{3(1-\chi)} + \frac{\chi^2\log\chi}{3(1-\chi)^2} - \frac{(2+\chi)\log(1-\chi)}{3\chi}
\\
\bar D_{1,2,1,2}(\chi) = &\ - \frac{1}{3\chi(1-\chi)} - \frac{(3-2\chi)\log\chi}{3(1-\chi)^2} - \frac{(1+2\chi)\log(1-\chi)}{3\chi^2}\,,
\\
\bar D_{2,2,2,2}(\chi) = &\
- \frac{2(  1- \chi + \chi^2)}{15\chi^2(1 - \chi)^2 } - \frac{(5 - 5 \chi + 2 \chi^2) \log(\chi)}{15(1-\chi)^3} - \frac{(2 + \chi+ 2 \chi^2) \log(1 - \chi)}{15 \chi^3}\,,
\\
\bar D_{2,2,3,3}(\chi) = &\   -\frac{24 - 48\chi - 5\chi^2 + 29\chi^3 - 18\chi^4}{210 \chi^2(1 - \chi)^3} 
- \frac{(12 + 6\chi + 8\chi^2 + 9\chi^3) \log(1 - \chi)}{105 \chi^3}\nonumber\\
 &\ + \frac{\chi^2(28 - 28\chi + 9\chi^2) \log(\chi)}{105 (1 - \chi)^4}\,,
\\
\bar D_{2,3,2,3}(\chi) = &\ -\frac{18 - 29\chi + 5\chi^2 + 48\chi^3 - 24\chi^4}{210\chi^3(1 - \chi)^3 } - \frac{(9 + 8\chi + 6\chi^2 + 12\chi^3) \log(1 - \chi)}{105 \chi^4} \nonumber \\
&\ + \frac{(-35 + 56\chi - 42\chi^2 + 12\chi^3) \log(\chi)}{105 (1 - \chi)^4}\,,
\\
\bar D_{3,3,3,3}(\chi) = &\ \frac{-24 + 72\chi - 74\chi^2 + 28\chi^3 - 74\chi^4 + 72\chi^5 - 24\chi^6}{315(1 - \chi)^4 \chi^4}\\
&\hspace{-1.4cm} - \frac{4(2 + \chi + \chi^2 + \chi^3 + 2\chi^4) \log(1 - \chi)}{105 \chi^5}
 - \frac{4(7 - 14\chi + 16\chi^2 - 9\chi^3 + 2\chi^4) \log(\chi)}{105 (1 - \chi)^5}\,,\nonumber
\end{align}
along with some others obtained from the following identities \cite{Dolan:2000ut}
\begin{align}
\bar D_{\Delta+1,\Delta+1,\Delta,\Delta}(\chi) = &\ \tfrac{1}{\chi^2}\bar D_{\Delta,\Delta,\Delta+1,\Delta+1}(\chi)\,, 
\\
\bar D_{\Delta,\Delta+1,\Delta+1,\Delta}(\chi) = &\ \tfrac{1}{(1-\chi)^2}\bar D_{\Delta,\Delta,\Delta+1,\Delta+1}(1-\chi)\,,
\\
\bar D_{\Delta+1,\Delta,\Delta,\Delta+1}(\chi) = &\ \bar D_{\Delta,\Delta,\Delta+1,\Delta+1}(1-\chi)\,,
\\
\bar D_{\Delta+1,1,\Delta+1,1}(\chi) = &\ \bar D_{1,\Delta+1,1,\Delta+1}(\chi)\,.
\end{align}

\bibliographystyle{JHEP}
\bibliography{ref.bib}

\providecommand{\href}[2]{#2}\begingroup\raggedright\begin{thebibliography}{10}

\bibitem{Cardy:1984bb}
J.~L. Cardy, \emph{{Conformal Invariance and Surface Critical Behavior}}, \href{https://doi.org/10.1016/0550-3213(84)90241-4}{\emph{Nucl. Phys. B} {\bfseries 240} (1984) 514}.

\bibitem{Cardy:2004hm}
J.~L. Cardy, \emph{{Boundary conformal field theory}},  \href{https://arxiv.org/abs/hep-th/0411189}{{\ttfamily hep-th/0411189}}.

\bibitem{Drukker:2006xg}
N.~Drukker and S.~Kawamoto, \emph{{Small deformations of supersymmetric Wilson loops and open spin-chains}}, \href{https://doi.org/10.1088/1126-6708/2006/07/024}{\emph{JHEP} {\bfseries 07} (2006) 024} [\href{https://arxiv.org/abs/hep-th/0604124}{{\ttfamily hep-th/0604124}}].

\bibitem{Giombi:2017cqn}
S.~Giombi, R.~Roiban and A.~A. Tseytlin, \emph{{Half-BPS Wilson loop and AdS$_2$/CFT$_1$}}, \href{https://doi.org/10.1016/j.nuclphysb.2017.07.004}{\emph{Nucl. Phys. B} {\bfseries 922} (2017) 499} [\href{https://arxiv.org/abs/1706.00756}{{\ttfamily 1706.00756}}].

\bibitem{Billo:2016cpy}
M.~Bill\`o, V.~Gon\c{c}alves, E.~Lauria and M.~Meineri, \emph{{Defects in conformal field theory}}, \href{https://doi.org/10.1007/JHEP04(2016)091}{\emph{JHEP} {\bfseries 04} (2016) 091} [\href{https://arxiv.org/abs/1601.02883}{{\ttfamily 1601.02883}}].

\bibitem{Gaiotto:2014kfa}
D.~Gaiotto, A.~Kapustin, N.~Seiberg and B.~Willett, \emph{{Generalized Global Symmetries}}, \href{https://doi.org/10.1007/JHEP02(2015)172}{\emph{JHEP} {\bfseries 02} (2015) 172} [\href{https://arxiv.org/abs/1412.5148}{{\ttfamily 1412.5148}}].

\bibitem{Maldacena:1998im}
J.~M. Maldacena, \emph{{Wilson loops in large N field theories}}, \href{https://doi.org/10.1103/PhysRevLett.80.4859}{\emph{Phys. Rev. Lett.} {\bfseries 80} (1998) 4859} [\href{https://arxiv.org/abs/hep-th/9803002}{{\ttfamily hep-th/9803002}}].

\bibitem{Correa:2012at}
D.~Correa, J.~Henn, J.~Maldacena and A.~Sever, \emph{{An exact formula for the radiation of a moving quark in N=4 super Yang Mills}}, \href{https://doi.org/10.1007/JHEP06(2012)048}{\emph{JHEP} {\bfseries 06} (2012) 048} [\href{https://arxiv.org/abs/1202.4455}{{\ttfamily 1202.4455}}].

\bibitem{Maldacena:1997re}
J.~M. Maldacena, \emph{{The Large N limit of superconformal field theories and supergravity}}, \href{https://doi.org/10.4310/ATMP.1998.v2.n2.a1}{\emph{Adv. Theor. Math. Phys.} {\bfseries 2} (1998) 231} [\href{https://arxiv.org/abs/hep-th/9711200}{{\ttfamily hep-th/9711200}}].

\bibitem{Drukker:2012de}
N.~Drukker, \emph{{Integrable Wilson loops}}, \href{https://doi.org/10.1007/JHEP10(2013)135}{\emph{JHEP} {\bfseries 10} (2013) 135} [\href{https://arxiv.org/abs/1203.1617}{{\ttfamily 1203.1617}}].

\bibitem{Correa:2012hh}
D.~Correa, J.~Maldacena and A.~Sever, \emph{{The quark anti-quark potential and the cusp anomalous dimension from a TBA equation}}, \href{https://doi.org/10.1007/JHEP08(2012)134}{\emph{JHEP} {\bfseries 08} (2012) 134} [\href{https://arxiv.org/abs/1203.1913}{{\ttfamily 1203.1913}}].

\bibitem{Giombi:2018qox}
S.~Giombi and S.~Komatsu, \emph{{Exact Correlators on the Wilson Loop in $\mathcal{N}=4$ SYM: Localization, Defect CFT, and Integrability}}, \href{https://doi.org/10.1007/JHEP05(2018)109}{\emph{JHEP} {\bfseries 05} (2018) 109} [\href{https://arxiv.org/abs/1802.05201}{{\ttfamily 1802.05201}}].

\bibitem{Correa:2023lsm}
D.~H. Correa, V.~I. Giraldo-Rivera and M.~Lagares, \emph{{Integrable Wilson loops in ABJM: a Y-system computation of the cusp anomalous dimension}}, \href{https://doi.org/10.1007/JHEP06(2023)179}{\emph{JHEP} {\bfseries 06} (2023) 179} [\href{https://arxiv.org/abs/2304.01924}{{\ttfamily 2304.01924}}].

\bibitem{Pestun:2007rz}
V.~Pestun, \emph{{Localization of gauge theory on a four-sphere and supersymmetric Wilson loops}}, \href{https://doi.org/10.1007/s00220-012-1485-0}{\emph{Commun. Math. Phys.} {\bfseries 313} (2012) 71} [\href{https://arxiv.org/abs/0712.2824}{{\ttfamily 0712.2824}}].

\bibitem{Liendo:2018ukf}
P.~Liendo, C.~Meneghelli and V.~Mitev, \emph{{Bootstrapping the half-BPS line defect}}, \href{https://doi.org/10.1007/JHEP10(2018)077}{\emph{JHEP} {\bfseries 10} (2018) 077} [\href{https://arxiv.org/abs/1806.01862}{{\ttfamily 1806.01862}}].

\bibitem{Bianchi:2020hsz}
L.~Bianchi, G.~Bliard, V.~Forini, L.~Griguolo and D.~Seminara, \emph{{Analytic bootstrap and Witten diagrams for the ABJM Wilson line as defect CFT$_{1}$}}, \href{https://doi.org/10.1007/JHEP08(2020)143}{\emph{JHEP} {\bfseries 08} (2020) 143} [\href{https://arxiv.org/abs/2004.07849}{{\ttfamily 2004.07849}}].

\bibitem{Ferrero:2021bsb}
P.~Ferrero and C.~Meneghelli, \emph{{Bootstrapping the half-BPS line defect CFT in N=4 supersymmetric Yang-Mills theory at strong coupling}}, \href{https://doi.org/10.1103/PhysRevD.104.L081703}{\emph{Phys. Rev. D} {\bfseries 104} (2021) L081703} [\href{https://arxiv.org/abs/2103.10440}{{\ttfamily 2103.10440}}].

\bibitem{Ferrero:2023gnu}
P.~Ferrero and C.~Meneghelli, \emph{{Unmixing the Wilson line defect CFT. Part II. Analytic bootstrap}}, \href{https://doi.org/10.1007/JHEP06(2024)010}{\emph{JHEP} {\bfseries 06} (2024) 010} [\href{https://arxiv.org/abs/2312.12551}{{\ttfamily 2312.12551}}].

\bibitem{Ferrero:2023znz}
P.~Ferrero and C.~Meneghelli, \emph{{Unmixing the Wilson line defect CFT. Part I. Spectrum and kinematics}}, \href{https://doi.org/10.1007/JHEP05(2024)090}{\emph{JHEP} {\bfseries 05} (2024) 090} [\href{https://arxiv.org/abs/2312.12550}{{\ttfamily 2312.12550}}].

\bibitem{Rattazzi:2008pe}
R.~Rattazzi, V.~S. Rychkov, E.~Tonni and A.~Vichi, \emph{{Bounding scalar operator dimensions in 4D CFT}}, \href{https://doi.org/10.1088/1126-6708/2008/12/031}{\emph{JHEP} {\bfseries 12} (2008) 031} [\href{https://arxiv.org/abs/0807.0004}{{\ttfamily 0807.0004}}].

\bibitem{El-Showk:2012cjh}
S.~El-Showk, M.~F. Paulos, D.~Poland, S.~Rychkov, D.~Simmons-Duffin and A.~Vichi, \emph{{Solving the 3D Ising Model with the Conformal Bootstrap}}, \href{https://doi.org/10.1103/PhysRevD.86.025022}{\emph{Phys. Rev. D} {\bfseries 86} (2012) 025022} [\href{https://arxiv.org/abs/1203.6064}{{\ttfamily 1203.6064}}].

\bibitem{Correa:2021sky}
D.~H. Correa, V.~I. Giraldo-Rivera and M.~Lagares, \emph{{On the abundance of supersymmetric strings in AdS$_{3}$ \texttimes{} S $^{3}$ \texttimes{} S $^{3}$ \texttimes{} S $^{1}$ describing BPS line operators}}, \href{https://doi.org/10.1088/1751-8121/ac354d}{\emph{J. Phys. A} {\bfseries 54} (2021) 505401} [\href{https://arxiv.org/abs/2108.09380}{{\ttfamily 2108.09380}}].

\bibitem{Pozzi:2024xnu}
R.~G. Pozzi and D.~Trancanelli, \emph{{Bootstrap of the defect 1/2 BPS Wilson lines in N=4 Chern-Simons-matter theories}}, \href{https://doi.org/10.1103/PhysRevD.110.066006}{\emph{Phys. Rev. D} {\bfseries 110} (2024) 066006} [\href{https://arxiv.org/abs/2406.13571}{{\ttfamily 2406.13571}}].

\bibitem{Gimenez-Grau:2019hez}
A.~Gimenez-Grau and P.~Liendo, \emph{{Bootstrapping line defects in $\mathcal{N}=2$ theories}}, \href{https://doi.org/10.1007/JHEP03(2020)121}{\emph{JHEP} {\bfseries 03} (2020) 121} [\href{https://arxiv.org/abs/1907.04345}{{\ttfamily 1907.04345}}].

\bibitem{Borsato:2013qpa}
R.~Borsato, O.~Ohlsson~Sax, A.~Sfondrini, B.~Stefa\'nski and A.~Torrielli, \emph{{The all-loop integrable spin-chain for strings on AdS$_3 \times S^3 \times T^4$: the massive sector}}, \href{https://doi.org/10.1007/JHEP08(2013)043}{\emph{JHEP} {\bfseries 08} (2013) 043} [\href{https://arxiv.org/abs/1303.5995}{{\ttfamily 1303.5995}}].

\bibitem{Beisert:2006qh}
N.~Beisert, \emph{{The Analytic Bethe Ansatz for a Chain with Centrally Extended su(2|2) Symmetry}}, \href{https://doi.org/10.1088/1742-5468/2007/01/P01017}{\emph{J. Stat. Mech.} {\bfseries 0701} (2007) P01017} [\href{https://arxiv.org/abs/nlin/0610017}{{\ttfamily nlin/0610017}}].

\bibitem{Agmon:2020pde}
N.~B. Agmon and Y.~Wang, \emph{{Classifying Superconformal Defects in Diverse Dimensions Part I: Superconformal Lines}},  \href{https://arxiv.org/abs/2009.06650}{{\ttfamily 2009.06650}}.

\bibitem{McAvity:1993ue}
D.~M. McAvity and H.~Osborn, \emph{{Energy momentum tensor in conformal field theories near a boundary}}, \href{https://doi.org/10.1016/0550-3213(93)90005-A}{\emph{Nucl. Phys. B} {\bfseries 406} (1993) 655} [\href{https://arxiv.org/abs/hep-th/9302068}{{\ttfamily hep-th/9302068}}].

\bibitem{Dolan:2004mu}
F.~A. Dolan, L.~Gallot and E.~Sokatchev, \emph{{On four-point functions of 1/2-BPS operators in general dimensions}}, \href{https://doi.org/10.1088/1126-6708/2004/09/056}{\emph{JHEP} {\bfseries 09} (2004) 056} [\href{https://arxiv.org/abs/hep-th/0405180}{{\ttfamily hep-th/0405180}}].

\bibitem{Baume:2019aid}
F.~Baume, M.~Fuchs and C.~Lawrie, \emph{{Superconformal Blocks for Mixed 1/2-BPS Correlators with $SU(2)$ R-symmetry}}, \href{https://doi.org/10.1007/JHEP11(2019)164}{\emph{JHEP} {\bfseries 11} (2019) 164} [\href{https://arxiv.org/abs/1908.02768}{{\ttfamily 1908.02768}}].

\bibitem{Bliard:2024und}
G.~Bliard, \emph{{On multipoint Ward identities for superconformal line defects}},  \href{https://arxiv.org/abs/2405.15846}{{\ttfamily 2405.15846}}.

\bibitem{Eden:2001ec}
B.~Eden and E.~Sokatchev, \emph{{On the OPE of 1/2 BPS short operators in N=4 SCFT(4)}}, \href{https://doi.org/10.1016/S0550-3213(01)00492-8}{\emph{Nucl. Phys. B} {\bfseries 618} (2001) 259} [\href{https://arxiv.org/abs/hep-th/0106249}{{\ttfamily hep-th/0106249}}].

\bibitem{Arutyunov:2001qw}
G.~Arutyunov, B.~Eden and E.~Sokatchev, \emph{{On nonrenormalization and OPE in superconformal field theories}}, \href{https://doi.org/10.1016/S0550-3213(01)00529-6}{\emph{Nucl. Phys. B} {\bfseries 619} (2001) 359} [\href{https://arxiv.org/abs/hep-th/0105254}{{\ttfamily hep-th/0105254}}].

\bibitem{Eden:2001wg}
B.~Eden, S.~Ferrara and E.~Sokatchev, \emph{{(2,0) superconformal OPEs in D = 6, selection rules and nonrenormalization theorems}}, \href{https://doi.org/10.1088/1126-6708/2001/11/020}{\emph{JHEP} {\bfseries 11} (2001) 020} [\href{https://arxiv.org/abs/hep-th/0107084}{{\ttfamily hep-th/0107084}}].

\bibitem{Ferrara:2001uj}
S.~Ferrara and E.~Sokatchev, \emph{{Universal properties of superconformal OPEs for 1/2 BPS operators in 3 \ensuremath{<}= D \ensuremath{<}= 6}}, \href{https://doi.org/10.1088/1367-2630/4/1/302}{\emph{New J. Phys.} {\bfseries 4} (2002) 2} [\href{https://arxiv.org/abs/hep-th/0110174}{{\ttfamily hep-th/0110174}}].

\bibitem{Bliard:2023zpe}
G.~J.~S. Bliard, \emph{{Perturbative and non-perturbative analysis of defect correlators in AdS/CFT}}, Ph.D. thesis, Humboldt U., Berlin, 2023.
\newblock \href{https://arxiv.org/abs/2310.18137}{{\ttfamily 2310.18137}}.
\newblock 10.18452/27559.

\bibitem{Ferrero:2019luz}
P.~Ferrero, K.~Ghosh, A.~Sinha and A.~Zahed, \emph{{Crossing symmetry, transcendentality and the Regge behaviour of 1d CFTs}}, \href{https://doi.org/10.1007/JHEP07(2020)170}{\emph{JHEP} {\bfseries 07} (2020) 170} [\href{https://arxiv.org/abs/1911.12388}{{\ttfamily 1911.12388}}].

\bibitem{Remiddi:1999ew}
E.~Remiddi and J.~A.~M. Vermaseren, \emph{{Harmonic polylogarithms}}, \href{https://doi.org/10.1142/S0217751X00000367}{\emph{Int. J. Mod. Phys. A} {\bfseries 15} (2000) 725} [\href{https://arxiv.org/abs/hep-ph/9905237}{{\ttfamily hep-ph/9905237}}].

\bibitem{Alday:2015eya}
L.~F. Alday, A.~Bissi and T.~Lukowski, \emph{{Large spin systematics in CFT}}, \href{https://doi.org/10.1007/JHEP11(2015)101}{\emph{JHEP} {\bfseries 11} (2015) 101} [\href{https://arxiv.org/abs/1502.07707}{{\ttfamily 1502.07707}}].

\bibitem{Cavaglia:2023mmu}
A.~Cavagli\`a, N.~Gromov and M.~Preti, \emph{{Computing Four-Point Functions with Integrability, Bootstrap and Parity Symmetry}},  \href{https://arxiv.org/abs/2312.11604}{{\ttfamily 2312.11604}}.

\bibitem{Heemskerk:2009pn}
I.~Heemskerk, J.~Penedones, J.~Polchinski and J.~Sully, \emph{{Holography from Conformal Field Theory}}, \href{https://doi.org/10.1088/1126-6708/2009/10/079}{\emph{JHEP} {\bfseries 10} (2009) 079} [\href{https://arxiv.org/abs/0907.0151}{{\ttfamily 0907.0151}}].

\bibitem{Fitzpatrick:2010zm}
A.~L. Fitzpatrick, E.~Katz, D.~Poland and D.~Simmons-Duffin, \emph{{Effective Conformal Theory and the Flat-Space Limit of AdS}}, \href{https://doi.org/10.1007/JHEP07(2011)023}{\emph{JHEP} {\bfseries 07} (2011) 023} [\href{https://arxiv.org/abs/1007.2412}{{\ttfamily 1007.2412}}].

\bibitem{Gauntlett:1998kc}
J.~P. Gauntlett, R.~C. Myers and P.~K. Townsend, \emph{{Supersymmetry of rotating branes}}, \href{https://doi.org/10.1103/PhysRevD.59.025001}{\emph{Phys. Rev. D} {\bfseries 59} (1998) 025001} [\href{https://arxiv.org/abs/hep-th/9809065}{{\ttfamily hep-th/9809065}}].

\bibitem{Babichenko:2009dk}
A.~Babichenko, B.~Stefanski, Jr. and K.~Zarembo, \emph{{Integrability and the AdS(3)/CFT(2) correspondence}}, \href{https://doi.org/10.1007/JHEP03(2010)058}{\emph{JHEP} {\bfseries 03} (2010) 058} [\href{https://arxiv.org/abs/0912.1723}{{\ttfamily 0912.1723}}].

\bibitem{Drukker:2000ep}
N.~Drukker, D.~J. Gross and A.~A. Tseytlin, \emph{{Green-Schwarz string in AdS(5) x S**5: Semiclassical partition function}}, \href{https://doi.org/10.1088/1126-6708/2000/04/021}{\emph{JHEP} {\bfseries 04} (2000) 021} [\href{https://arxiv.org/abs/hep-th/0001204}{{\ttfamily hep-th/0001204}}].

\bibitem{Forini:2015mca}
V.~Forini, V.~G.~M. Puletti, L.~Griguolo, D.~Seminara and E.~Vescovi, \emph{{Remarks on the geometrical properties of semiclassically quantized strings}}, \href{https://doi.org/10.1088/1751-8113/48/47/475401}{\emph{J. Phys. A} {\bfseries 48} (2015) 475401} [\href{https://arxiv.org/abs/1507.01883}{{\ttfamily 1507.01883}}].

\bibitem{Pajer:2021bfr}
D.~Pajer, \emph{{On quantum corrections to BPS Wilson loops in superstring theory on $\mathbf{AdS_3\times S^3 \times T^4}$ with mixed flux}},  \href{https://arxiv.org/abs/2109.11318}{{\ttfamily 2109.11318}}.

\bibitem{Correa:2023thy}
D.~H. Correa, M.~G. Ferro and V.~I. Giraldo-Rivera, \emph{{Mixed boundary conditions in AdS$_{2}$/CFT$_{1}$ from the coupling with a Kalb-Ramond field}}, \href{https://doi.org/10.1007/JHEP02(2024)141}{\emph{JHEP} {\bfseries 02} (2024) 141} [\href{https://arxiv.org/abs/2312.13258}{{\ttfamily 2312.13258}}].

\bibitem{DHoker:1999mqo}
E.~D'Hoker, D.~Z. Freedman and L.~Rastelli, \emph{{AdS / CFT four point functions: How to succeed at z integrals without really trying}}, \href{https://doi.org/10.1016/S0550-3213(99)00526-X}{\emph{Nucl. Phys. B} {\bfseries 562} (1999) 395} [\href{https://arxiv.org/abs/hep-th/9905049}{{\ttfamily hep-th/9905049}}].

\bibitem{DHoker:1999kzh}
E.~D'Hoker, D.~Z. Freedman, S.~D. Mathur, A.~Matusis and L.~Rastelli, \emph{{Graviton exchange and complete four point functions in the AdS / CFT correspondence}}, \href{https://doi.org/10.1016/S0550-3213(99)00525-8}{\emph{Nucl. Phys. B} {\bfseries 562} (1999) 353} [\href{https://arxiv.org/abs/hep-th/9903196}{{\ttfamily hep-th/9903196}}].

\bibitem{Zhou:2018sfz}
X.~Zhou, \emph{{Recursion Relations in Witten Diagrams and Conformal Partial Waves}}, \href{https://doi.org/10.1007/JHEP05(2019)006}{\emph{JHEP} {\bfseries 05} (2019) 006} [\href{https://arxiv.org/abs/1812.01006}{{\ttfamily 1812.01006}}].

\bibitem{Bliard:2022xsm}
G.~Bliard, \emph{{Notes on n-point Witten diagrams in AdS$_{2}$}}, \href{https://doi.org/10.1088/1751-8121/ac7f6b}{\emph{J. Phys. A} {\bfseries 55} (2022) 325401} [\href{https://arxiv.org/abs/2204.01659}{{\ttfamily 2204.01659}}].

\bibitem{Carmi:2018qzm}
D.~Carmi, L.~Di~Pietro and S.~Komatsu, \emph{{A Study of Quantum Field Theories in AdS at Finite Coupling}}, \href{https://doi.org/10.1007/JHEP01(2019)200}{\emph{JHEP} {\bfseries 01} (2019) 200} [\href{https://arxiv.org/abs/1810.04185}{{\ttfamily 1810.04185}}].

\bibitem{Bianchi:2018scb}
L.~Bianchi, M.~Preti and E.~Vescovi, \emph{{Exact Bremsstrahlung functions in ABJM theory}}, \href{https://doi.org/10.1007/JHEP07(2018)060}{\emph{JHEP} {\bfseries 07} (2018) 060} [\href{https://arxiv.org/abs/1802.07726}{{\ttfamily 1802.07726}}].

\bibitem{Bianchi:2014laa}
M.~S. Bianchi, L.~Griguolo, M.~Leoni, S.~Penati and D.~Seminara, \emph{{BPS Wilson loops and Bremsstrahlung function in ABJ(M): a two loop analysis}}, \href{https://doi.org/10.1007/JHEP06(2014)123}{\emph{JHEP} {\bfseries 06} (2014) 123} [\href{https://arxiv.org/abs/1402.4128}{{\ttfamily 1402.4128}}].

\bibitem{Bianchi:2017ozk}
L.~Bianchi, L.~Griguolo, M.~Preti and D.~Seminara, \emph{{Wilson lines as superconformal defects in ABJM theory: a formula for the emitted radiation}}, \href{https://doi.org/10.1007/JHEP10(2017)050}{\emph{JHEP} {\bfseries 10} (2017) 050} [\href{https://arxiv.org/abs/1706.06590}{{\ttfamily 1706.06590}}].

\bibitem{Correa:2014aga}
D.~H. Correa, J.~Aguilera-Damia and G.~A. Silva, \emph{{Strings in $AdS_4 \times \mathbb{CP}^{3}$ Wilson loops in $\mathcal N=$6 super Chern-Simons-matter and bremsstrahlung functions}}, \href{https://doi.org/10.1007/JHEP06(2014)139}{\emph{JHEP} {\bfseries 06} (2014) 139} [\href{https://arxiv.org/abs/1405.1396}{{\ttfamily 1405.1396}}].

\bibitem{Drukker:2008zx}
N.~Drukker, J.~Plefka and D.~Young, \emph{{Wilson loops in 3-dimensional N=6 supersymmetric Chern-Simons Theory and their string theory duals}}, \href{https://doi.org/10.1088/1126-6708/2008/11/019}{\emph{JHEP} {\bfseries 11} (2008) 019} [\href{https://arxiv.org/abs/0809.2787}{{\ttfamily 0809.2787}}].

\bibitem{Cornagliotto:2017dup}
M.~Cornagliotto, M.~Lemos and V.~Schomerus, \emph{{Long Multiplet Bootstrap}}, \href{https://doi.org/10.1007/JHEP10(2017)119}{\emph{JHEP} {\bfseries 10} (2017) 119} [\href{https://arxiv.org/abs/1702.05101}{{\ttfamily 1702.05101}}].

\bibitem{OhlssonSax:2014jtq}
O.~Ohlsson~Sax, A.~Sfondrini and B.~Stefanski, \emph{{Integrability and the Conformal Field Theory of the Higgs branch}}, \href{https://doi.org/10.1007/JHEP06(2015)103}{\emph{JHEP} {\bfseries 06} (2015) 103} [\href{https://arxiv.org/abs/1411.3676}{{\ttfamily 1411.3676}}].

\bibitem{Drukker:2022pxk}
N.~Drukker, Z.~Kong and G.~Sakkas, \emph{{Broken Global Symmetries and Defect Conformal Manifolds}}, \href{https://doi.org/10.1103/PhysRevLett.129.201603}{\emph{Phys. Rev. Lett.} {\bfseries 129} (2022) 201603} [\href{https://arxiv.org/abs/2203.17157}{{\ttfamily 2203.17157}}].

\bibitem{Barrat:2021tpn}
J.~Barrat, P.~Liendo, G.~Peveri and J.~Plefka, \emph{{Multipoint correlators on the supersymmetric Wilson line defect CFT}}, \href{https://doi.org/10.1007/JHEP08(2022)067}{\emph{JHEP} {\bfseries 08} (2022) 067} [\href{https://arxiv.org/abs/2112.10780}{{\ttfamily 2112.10780}}].

\bibitem{Giombi:2023zte}
S.~Giombi, S.~Komatsu, B.~Offertaler and J.~Shan, \emph{{Boundary reparametrizations and six-point functions on the AdS$_{2}$ string}}, \href{https://doi.org/10.1007/JHEP08(2024)196}{\emph{JHEP} {\bfseries 08} (2024) 196} [\href{https://arxiv.org/abs/2308.10775}{{\ttfamily 2308.10775}}].

\bibitem{Rosenhaus:2018zqn}
V.~Rosenhaus, \emph{{Multipoint Conformal Blocks in the Comb Channel}}, \href{https://doi.org/10.1007/JHEP02(2019)142}{\emph{JHEP} {\bfseries 02} (2019) 142} [\href{https://arxiv.org/abs/1810.03244}{{\ttfamily 1810.03244}}].

\bibitem{Freedman:1998tz}
D.~Z. Freedman, S.~D. Mathur, A.~Matusis and L.~Rastelli, \emph{{Correlation functions in the CFT(d) / AdS(d+1) correspondence}}, \href{https://doi.org/10.1016/S0550-3213(99)00053-X}{\emph{Nucl. Phys. B} {\bfseries 546} (1999) 96} [\href{https://arxiv.org/abs/hep-th/9804058}{{\ttfamily hep-th/9804058}}].

\bibitem{Fitzpatrick:2011ia}
A.~L. Fitzpatrick, J.~Kaplan, J.~Penedones, S.~Raju and B.~C. van Rees, \emph{{A Natural Language for AdS/CFT Correlators}}, \href{https://doi.org/10.1007/JHEP11(2011)095}{\emph{JHEP} {\bfseries 11} (2011) 095} [\href{https://arxiv.org/abs/1107.1499}{{\ttfamily 1107.1499}}].

\bibitem{Penedones:2010ue}
J.~Penedones, \emph{{Writing CFT correlation functions as AdS scattering amplitudes}}, \href{https://doi.org/10.1007/JHEP03(2011)025}{\emph{JHEP} {\bfseries 03} (2011) 025} [\href{https://arxiv.org/abs/1011.1485}{{\ttfamily 1011.1485}}].

\bibitem{Dolan:2000ut}
F.~A. Dolan and H.~Osborn, \emph{{Conformal four point functions and the operator product expansion}}, \href{https://doi.org/10.1016/S0550-3213(01)00013-X}{\emph{Nucl. Phys. B} {\bfseries 599} (2001) 459} [\href{https://arxiv.org/abs/hep-th/0011040}{{\ttfamily hep-th/0011040}}].

\end{thebibliography}\endgroup

\end{document}